\documentclass[fleqn,usenatbib]{mnras}

\usepackage{newtxtext,newtxmath}

\usepackage[T1]{fontenc}
\usepackage{ae,aecompl}

\usepackage{graphicx}	
\usepackage{amsmath}	
\usepackage{amssymb}	

\title[NGC 1172 and its peculiar globular cluster system]{Wide-field study of the peculiar globular cluster system hosted by the field lenticular NGC\,1172}

\author[A. I. Ennis et al.]{
Ana In\'es Ennis$^{1,2}$\thanks{E-mail: ennis.ana@gmail.com}, Juan Pablo Caso$^{1,2}$, Lilia P. Bassino$^{1,2}$, Ricardo Salinas$^{3}$ 
\newauthor{and Mat\'ias G\'omez$^{4}$} \\
$^{1}$Facultad de Ciencias Astron\'omicas y Geof\'isicas de la Universidad Nacional de La Plata, and Instituto de Astrof\'isica de La Plata\\ (CCT La Plata -- CONICET, UNLP), Paseo del Bosque S/N, B1900FWA La Plata, Argentina\\
$^{2}$Consejo Nacional de Investigaciones Cient\'ificas y T\'ecnicas, Godoy Cruz 2290, C1425FQB, Ciudad Aut\'onoma de Buenos Aires, Argentina\\
$^{3}$Gemini Observatory, Casilla 603, La Serena, Chile\\
$^{4}$Departamento de Ciencias F\'isicas, Facultad de Ciencias Exactas, Universidad Andres Bello, Santiago, Chile\\
}

\date{Accepted XXX. Received YYY; in original form ZZZ}

\pubyear{2020}

\begin{document}
\label{firstpage}
\pagerange{\pageref{firstpage}--\pageref{lastpage}}
\maketitle

\begin{abstract}
We present a wide-field study of the globular cluster system (GCS) of the field lenticular galaxy NGC\,1172, based on observations from GMOS/Gemini (optical), FourStar/Magellan (NIR), and archival data from ACS/\textit{HST} (optical). This analysis covers the full extension of the GCS, and results in a value of specific frequency ($S_{N}=8.6\pm1.5$) peculiarly high for an intermediate-mass galaxy in a low-density environment such as this one. We find that the GCS appears to be bimodal, although the colour distribution is narrow and does not allow for an accurate separation of the subpopulations. However, the combination of optical and NIR filters allows us to obtain an estimation of the metallicity distribution based on the photometry, which supports bimodality. We conclude that the presence of a large fraction of metal-poor globular clusters (GCs) and the high specific frequency point to NGC\,1172 having accreted a significant amount of GCs from low-mass satellites in the past. 
\end{abstract}

\begin{keywords}
galaxies: elliptical and lenticular, cD -- galaxies: star clusters: individual: NGC 1172 -- galaxies: evolution
\end{keywords}



\section{Introduction}
\label{intro.sec}

The properties of a globular cluster system (GCS) often unveil pieces of the evolutionary history of the galaxy that hosts it \citep[e.g][]{kruijssen2019}. Imprints of the processes the host galaxy has undergone can be found in properties of the system, such as alterations in its spatial distribution hinting at recent interactions, or the presence of intermediate-age or younger globular clusters (GCs) indicating accretion episodes \citep[e.g.][]{sesto2018,caso2019a}. 

The majority of the GC population in early-type galaxies (ETGs) consists of old objects \citep[e.g.][]{fahrion2020a,fahrion2020b}. The lack of a spread in age allows for colour distributions to be connected with metallicity distributions. In most bright galaxies, the colour distribution of a GCS is bimodal, and this is taken to indicate the presence of two subpopulations with different metallicities, a `blue' (more metal poor) and a `red' (more metal rich) one \citep{peng2006}.  This is supported by these subpopulations presenting different behaviours, in some cases in their spatial distribution, and where spectroscopic analysis was possible, differences in their kinematics were also detected \citep{pota2013}. Given the necessary  high resolution required to obtain metallicities spectroscopically, there are only a few GCS for which metallicity distributions where obtained directly, but they have also shown bimodality \citep[e.g.][]{villaume2019}. However, the transformation from colour to metallicity has been questioned and proven to be non-linear, in addition to showing sensitivity to the modelling of the horizontal branch in GCs \citep{yoon2006,lee2019}. 

The richness of a GCS in relation to its host galaxy has been largely discussed in the literature in terms of the efficiency of the galaxy in forming and disrupting GCs \citep{liu2019,li2019}, and it has been associated with the luminosity, stellar mass and the total mass of the galaxy through different parameters \citep[e.g.][]{spitler2008,hudson2014}. One of the most classic of these parameters is the specific frequency \citep{harris1981}, defined as the number of GCs normalized to the luminosity of the galaxy \footnote{$S_{N}=N_{GC}10^{0.4(M_{V}+15)}$}. It quantifies the richness of the system in relation to the luminosity of its host galaxy, and tracing its variations across morphological types, environments and different galaxy masses has allowed for a better understanding of the assembly and evolution of GCS. More massive galaxies show larger values of $S_{N}$, whereas dwarf galaxies show a wider spread of values since they are more easily affected by their environment, either increasing their accretion rates or causing them to have most of their GCs stripped from them \citep{renaud2018}. 

NGC\,1172 is an intermediate-mass galaxy, with an absolute magnitude of $M_{V}=-19.3$ adopting a distance of $d=20.8\,\rm{Mpc}$ \citep{cho2012}. It is located in the field, with no bright neighbours in a radius of 180\,kpc. A study of its GCS was presented by \cite{cho2012}, by means of ACS observations, and they obtain a specific frequency $S_{N}=9.18\pm4.41$, which is strikingly large for a galaxy with an intermediate luminosity \citep{harris2013}. Though no other cases of galaxies with similar masses are found in the literature with such high values of $S_{N}$ it is worth noting that NGC\,1172 is located in a very low-density environment, where ETGs are usually found to be hosting poorly populated GCS \citep{caso2013,salinas2015}. Environment plays a key role in the formation and evolution of galaxies since it constrains the accreted mass in a very direct way. Galaxies in low-density environments are expected to have lower accretion rates than those in high-density environments, probably due to the fact that having less neighbours in their surroundings makes encounters less frequent. This leads to poor GCS, but also to bluer galaxies \citep{lacerna2016}, with lighter haloes \citep{niemi2010}. For this reason, most massive early-type galaxies (ETGs) are found in clusters or high density groups, hosting GCS which are largely populated and spatially extended \citep[][e.g.]{cantiello2020}, pointing to the connection between the build-up of the GCs and the accretion history of the galaxy.

The motivation of this work is to analyse the full extent of the GCS of NGC\,1172 using optical and NIR data from GMOS (Gemini South) and FourStar (Magellan/Baade telescope at Las Campanas Observatory) which provide us with larger fields of view (FOVs) than the one in ACS, in order to get an in-depth look at the properties of the system, and recalculate the specific frequency, looking for confirmation of the value in \citet{cho2012} and for possible hints that might help unveil the evolutionary processes that built this population of GCs. The paper is structured as follows. In Section 2 of this work we introduce the data and describe the data reduction process. In Section 3 we analyse the results, including the selection of the sample. In Section 4, we present the discussion of these results, summarized in Section 5.

\section{Observations and data reduction}
\subsection{Optical data}

The optical data was obtained with the Gemini Multi-Object Spectrograph (GMOS) camera on Gemini South (FOV of $5.5\arcmin\times 5.5\arcmin$), over the course of two programs (Program IDs: GS-2016B-Q37 and GS-2017B-Q38). The first program consisted of observations of a single field containing the galaxy in the $g'$, $r'$, $i'$ and $z'$ filters, whereas the second program observed two adjacent fields in order to ensure full spatial coverage of the GCS and a clean area to estimate background corrections, in the $g'$, $r'$ and $i'$ filters. In all cases, a binning of $2\times2$ was used, resulting in a scale of $0.146\,\rm{arcsec}\,\rm{pixel}^{-1}$. 

From the Gemini Observatory Archive, calibration images such as BIAS and FLAT-FIELDs were downloaded to apply corrections to the data, using tasks from the \textsc{iraf} Gemini package for the processing. In the case of the $z'$ filter, it was also necessary to use  images to correct the fringing patterns since their presence can decrease the accuracy and quality of the photometry.  For each of the fields, the set of observations were slightly dithered so that their final combination resulted in an image free of the gaps produced by both bad columns or pixels and the physicals gaps on the detector. Finally, using the task FMEDIAN, we modeled and removed the light from the galaxy with two successive filters of different sizes to facilitate the detection of point-like sources. Further analysis of artificial stars added to the images and reduced in the same manner have shown that this procedure does not affect the photometry of point sources.

A field containing standard stars (E2-A, \citealt{smith2002}) was also observed, so that the instrumental magnitudes could be converted to standard ones using Equation \ref{eq:std}, for which we calculated zero-points and colour terms (CT). The $k_{CP}$ coefficients correspond to the median atmospheric extinction at Cerro Pach\'on where Gemini South is located, and were obtained from the Gemini Observatory website \footnote{\url{http://http://www.gemini.edu/}}.

\begin{equation}
\begin{split}
m_{\rm std}= & m_{\rm zero}-2.5\,\log_{10}\left(N(e-)/exptime\right) \\
& -k_{CP}(airmass-1.0)+CT(colour)
\end{split}
\label{eq:std}
\end{equation}

\subsection{Near-infrared data}

The near-infrared data was obtained with the FourStar camera \citep[FOV of $10.8\arcmin\times 10.8\arcmin$,][]{persson2013} mounted on the Baade telescope in Las Campanas Observatory, in the $K_{s}$ filter, on September 12, 2017. Two fields were observed, both containing the galaxy. These observations were processed using THELI \citep{erben2005,schirmer2013}, and bias and flat field images taken in the same night were used to correct them. The background was modelled using the chips furthest from the galaxy and then substracted from all the observations before combining them, in order to allow for an accurate astrometry. For the sky model, all sources were masked using a mask expansion factor of $11$ \footnote{\url{https://www.astro.uni-bonn.de/theli/}} as to eliminate as much light from the galaxy halo as possible from the background model, and we used a dynamic model since the background was not stable throughout the night of the observation. The coaddition of all the background-subtracted exposures was performed using the astrometry from the objects from 2MASS catalogue \citep{skrutskie2006} found in the field, which were also used for calibrating the photometry. 

The total exposure time on target considering both pointings was of $11385\,s$, divided into 495 expositions of $11.5\,s$ for each pointing.

\begin{figure}
    \centering
    \includegraphics[width=0.4\textwidth]{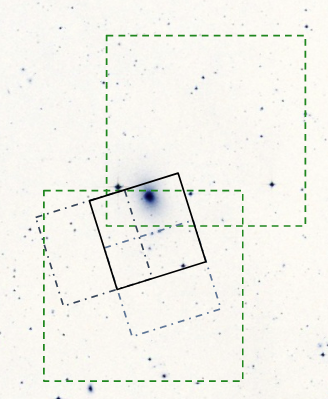}    
    \caption{DSS2 image of the region containing NGC\,1172, with the FourStar pointing shown in green dashed lines and the GMOS pointings shown in solid black lines (central field) and dashed-dotted blue lines (adjacent fields). North is up, East to the left, and the angular size of the image is $\sim 15\arcmin\times20\arcmin$.}
    \label{fig:fields}
\end{figure}

The superposition of all the observed fields for all instruments is shown in Figure \ref{fig:fields}.

\begin{table}
\centering
 \caption{Exposure times and effective wavelengths for each GMOS filter}
 \label{tab:exp}
 \begin{tabular}{|cccc|}
  \hline
  {Filter} & {$\lambda_{\rm eff}$ [nm]} & \multicolumn{2}{c|}{Exposure Time [s]}  \\
 &  & Central Field & Adjacent Fields \\

  \hline
  $g'$ & 475 & $8\times320$ & $8\times320$ \\[2pt] 
  $r'$ & 630 & $7\times150$ & $6\times150$ \\[2pt]
  $i'$ & 780 & $7\times120$ & $6\times120$ \\[2pt]
  $z'$ & 925 & $17\times180$ & - \\[2pt]
  \hline

\end{tabular}
\end{table}

\subsection{Detection of sources}
The same method was used on both sets of data. We used SE{\sc xtractor} \citep{bertin1996} to build the initial catalogue of sources, running it over every filter and then choosing the catalogue generated from the $i'$ filter,  since it was the most populated one. To identify point-like sources we made use of the $CLASS\_STAR$ parameter, which ranges from 0 (extended sources) to 1 (point sources). Our chosen criterion is $CLASS\_STAR < 0.5$, which excludes most galaxies that might be present in the background.

With DAOPHOT tasks within {\sc IRAF} \citep{stetson1987} we calculated aperture photometry magnitudes, and then used a set of bright stars relatively isolated and spanning the entire field of view in each case, to build the point source function (PSF) that corresponds to each filter and each field. With it, we can obtain accurate magnitudes using the allstar task. This task also runs statistical analysis that determine the goodness of the fit, allowing us to separate objects that are more extended or too dim to be accurately measured. These magnitudes were corrected for galactic extinction using the absorption coefficients obtained from \citet{schlafly2011}, with $E(B-V)=0.04$. 

\begin{figure}
    \centering
    \includegraphics[width=0.4\textwidth]{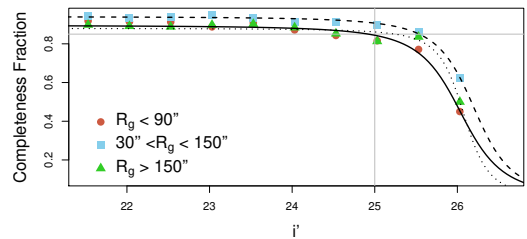}
    \caption{Completeness curves for three different regions of the GMOS fields. The gray lines show the limiting magnitude and the percentage reached.}
    \label{fig:compl}
\end{figure}

In addition, a completeness test was carried out adding 35000 sources of different magnitudes to the images in every filter, and then repeating the process described previously to perform photometry. We then counted the number of recovered sources across all filters for the central field, and the available ones for the adjacent ones, obtaining consistent completeness curves. From this, as shown in Figure \ref{fig:compl}, we obtain a level of completeness of 85\,per cent 
at $i'\sim25$.

\subsection{Additional \textit{HST} data}

In addition to our observations, we complemented our results using data observed with HST/ACS Wide Field Camera (WFC) in
$F475W$ and $F850LP$ (programme ID 10554), downloaded from the Mikulski Archive for Space Telescopes (MAST)\footnote{Based on
observations made with the NASA/ESA Hubble Space Telescope, obtained
from the data archive at the Space Telescope Science Institute. STScI
is operated by the Association of Universities for Research in Astronomy,
Inc. under NASA contract NAS 5-26555.}. The fields are centred on NGC\,1172 and IC\,2035. The latter is an elliptical galaxy with a sparse GCS, presenting around 50 members according with \citet{cho2012}, and the regions in its FOV at galactocentric distances larger than $\approx 1$\,arcmin were used to determine the contamination for the NGC\,1172 field. Both galaxies present intermediate galactic latitudes, but slightly different ($b\approx -57^{\circ}$ for NGC\,1172 and $b\approx-47^{\circ}$ for IC\,2035). In order to test these differences, the Besan\c{c}on Galactic models \citep{rob03} were used to build up simulated catalogues spanning $1\,{\rm deg^2}$ and centred on both galaxies. From the default parameters and the selection criteria $20 < z' < 25$\,mag and $0.6 < (g'-z') < 1.45$\,mag, the foreground contamination results $\approx 0.36$ and $\approx 0.47$\,objects arcmin$^{-2}$ for NGC\,1172 and IC\,2035, respectively, which implies $\approx 4-5$ objects for the ACS FOV. Hence, the foreground contamination in both fields is low enough to justify the use of the latter one as a comparison field. These observations were processed in a similar manner to the GMOS and FourStar data, subtracting the extended galaxy light using FMEDIAN and detecting sources using SE{\sc xtractor}. In this case, the selection of sources was based on constraints on the elongation and full width at half-maximum, following criteria taken from works which use similar instrumental setup for galaxies at a comparable distance \citep[e.g.][]{jordan2004,jordan2007}. In the case of ACS data, only aperture
photometry was performed and mean aperture corrections were applied, obtained from an analysis of the parameters using ISHAPE \citep[][]{larsen1999}. Completeness was calculated dividing the region in rings at different galactocentric distances, as to take into consideration differential effects in the radial distribution. Further information about the photometry of this dataset can be found in \citet{caso2019b}.

\section{Results}

\subsection{GC candidates selection}
\begin{figure}
    \centering
    \includegraphics[width=0.2\textwidth]{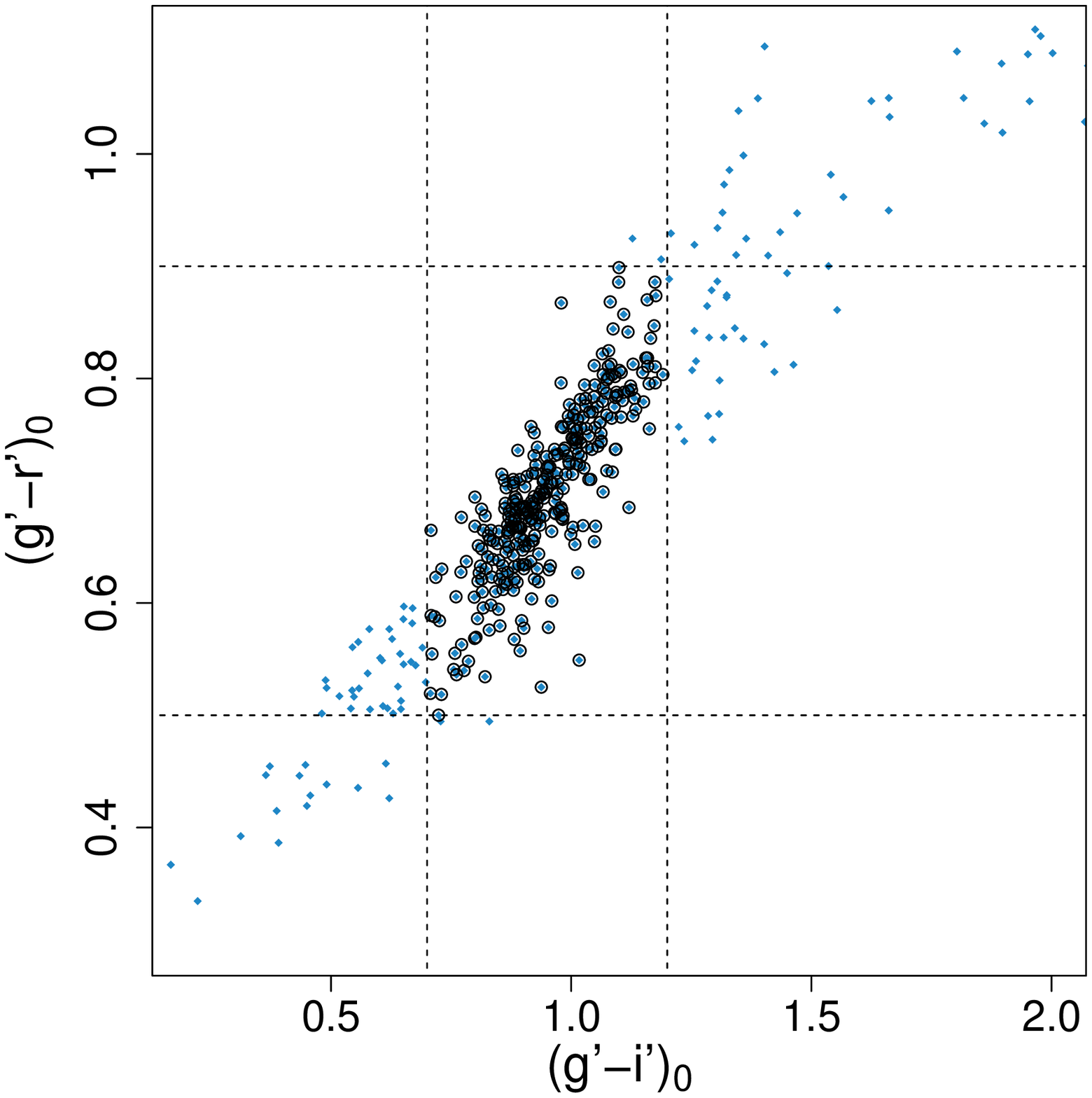}
    \includegraphics[width=0.2\textwidth]{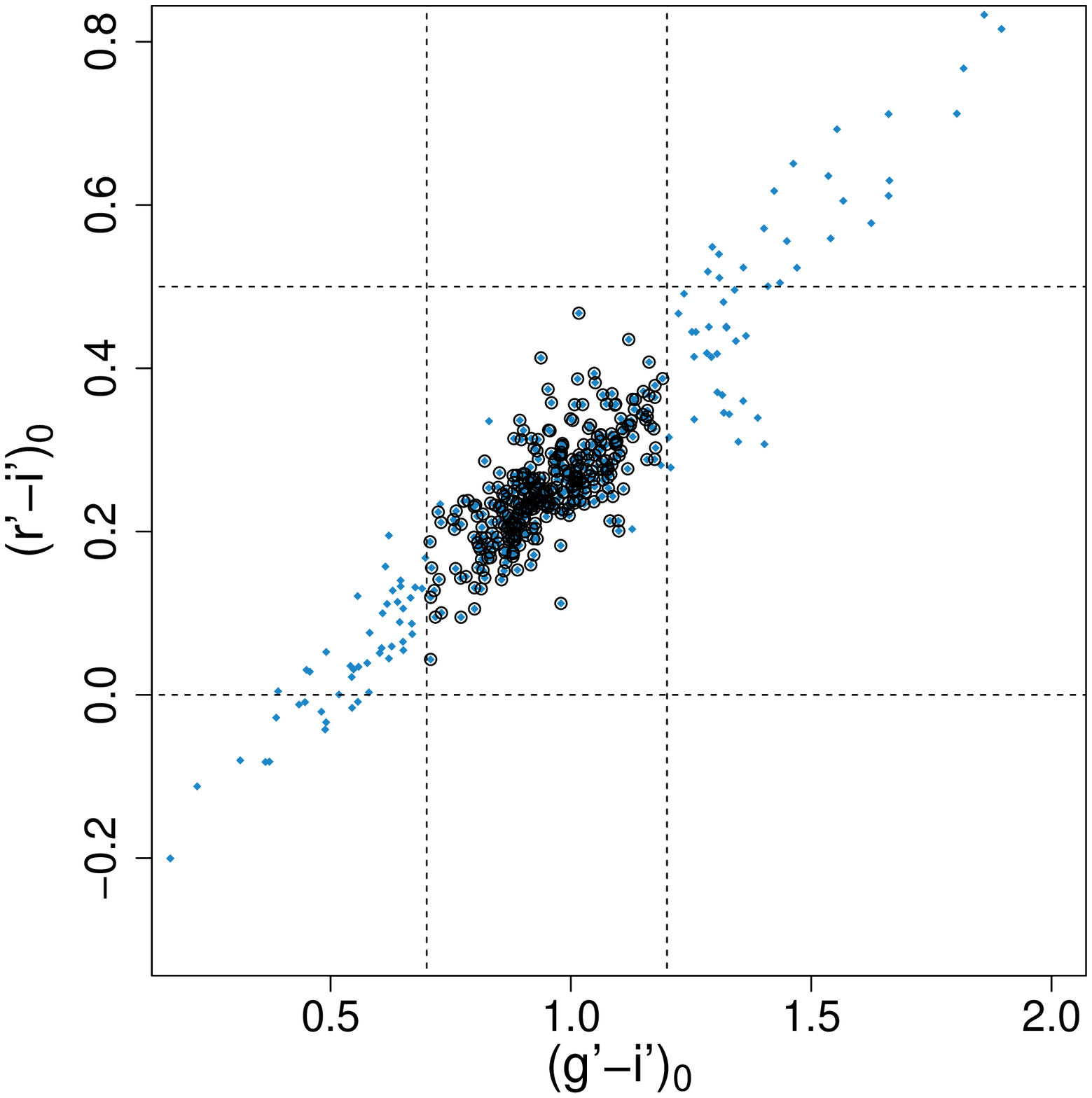}

    \includegraphics[width=0.2\textwidth]{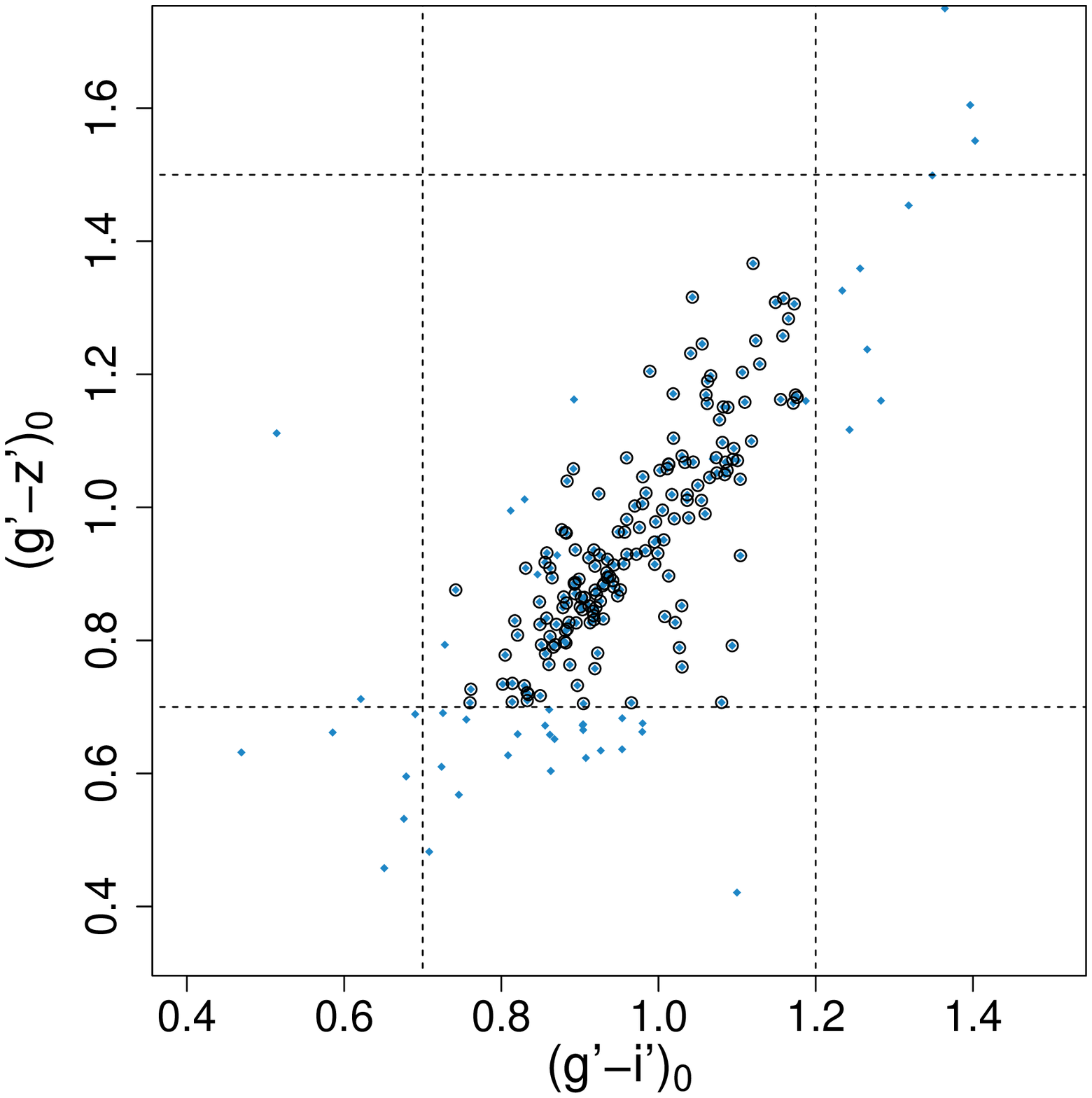}
    \includegraphics[width=0.2\textwidth]{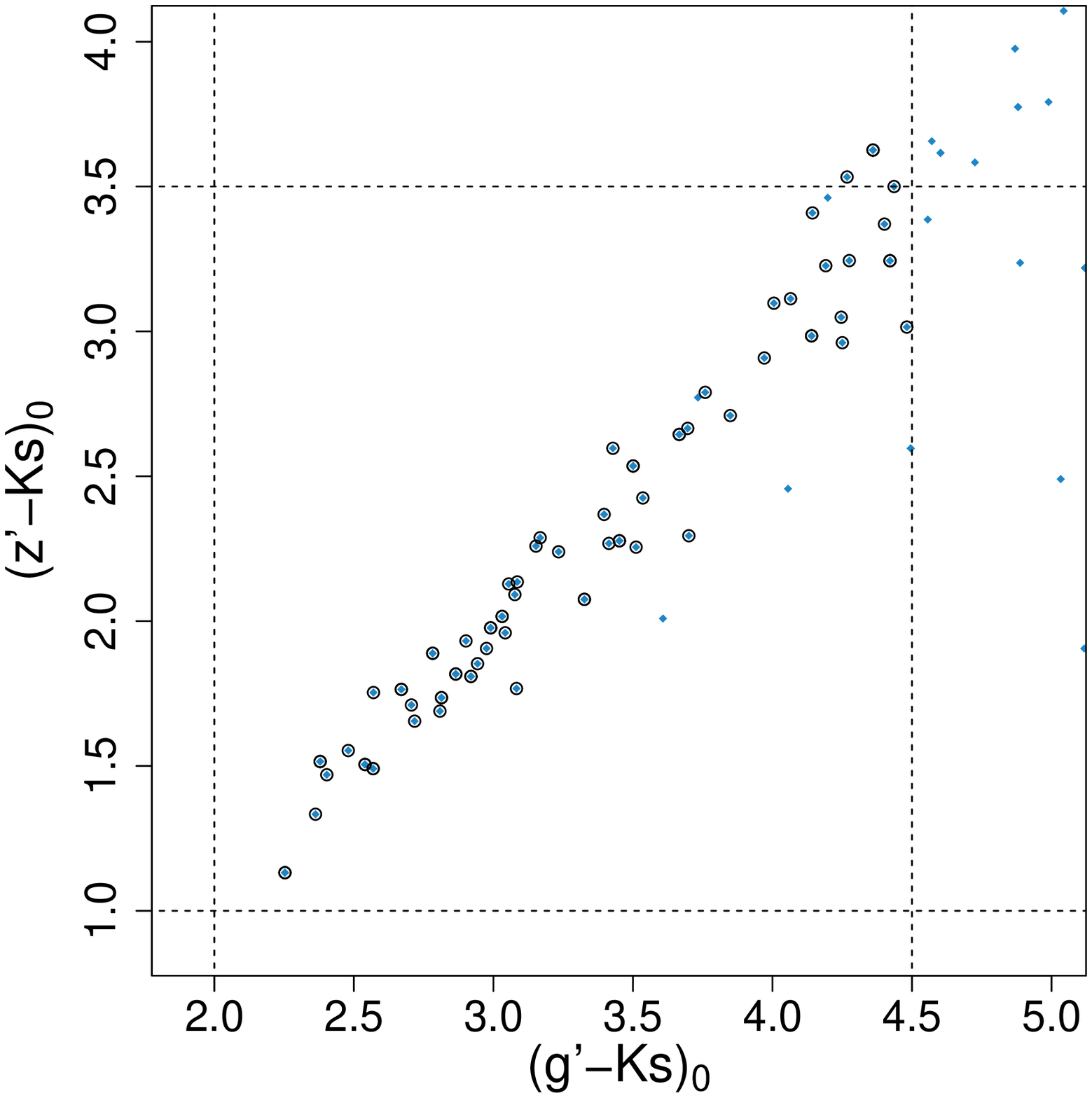}    
    \caption{Colour-colour distributions of the point like sources in the field brighter than $i'=25$, shown in smaller, light-blue diamonds. The black circles represent the GC candidates, as selected by the colour limits (shown in dotted lines).}
    \label{fig:colcol}
\end{figure}

Following \cite{bassino2017} in the case of $g'r'i'$ combinations, we applied colour limits to select the sample of GC candidates. We also conducted a visual inspection in the colours involving $z'$ and $Ks$ in order to set the limits, comparing with the ones used in \cite{chiessantos2012}. Fig. \ref{fig:colcol} shows some of the colour-colour diagrams with the obtained limits, each corresponding to the sample built from matching the different filters which resulted in different amounts of GC candidates in each case. It is worth mentioning that the last pannel shows the least amount of sources, a consequence of combining the $z'$ filter which has the smallest spatial coverage, and the $Ks$ filter, where the depth is significantly lower than in the GMOS filters. The chosen limits were $0.7<(g'-i')<1.2$, $0<(r'-i')<0.5$, $0.5<(g'-r')<0.9$, $0.7<(g'-z')<1.5$ and $2<(g'-Ks)<4.5$. It is important to note that for many objects presenting $g'r'i'$ magnitudes, either $z'$ or $Ks$ or both may be missing due to being located in the adjacent fields or to not being bright enough, respectively. In these cases, only the limits in the available colours were applied.

\begin{figure}
    \centering
    \includegraphics[width=0.3\textwidth]{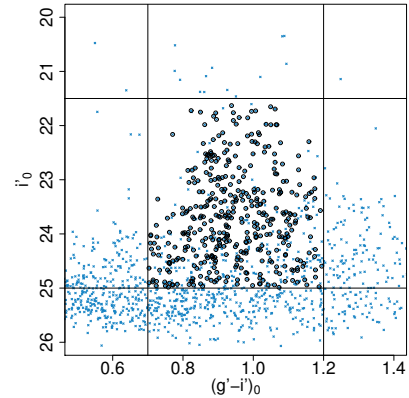}
    \caption{Colour-magnitude diagram of all the point like sources present in the 
    GMOS fields shown as blue crosses. The black circles represent the sample selected as GC candidates. The limits in magnitude (horizontal solid lines) correspond to a completeness level in the faint end, and to the removal of potential contaminants in the bright end. The limits in colour (vertical solid lines) were obtained examining the colour-colour distributions in Figure \ref{fig:colcol}.}
    \label{fig:colmag}
\end{figure}

In addition, we excluded objects with magnitudes brighter than $M_{V}=-11$ so as to clean the sample from potential contaminants, since the analysis of the luminosity function for old GCS in the literature shows that the probability of finding GCs at this luminosity is very low. \citep{harris2014}. The few objects within the colour boundaries that are above this limit might be ultra-compact dwarfs (UCDs), which will not be studied in this work. On the faint end, the limit in magnitude corresponds to the completeness level mentioned in the previous section. These limits can be seen in the colour-magnitude diagram shown in Fig. \ref{fig:colmag}.

\subsection{Colour distribution and subpopulations}
\label{dcol.sec}

\begin{figure}
    \centering
    \includegraphics[width=0.3\textwidth]{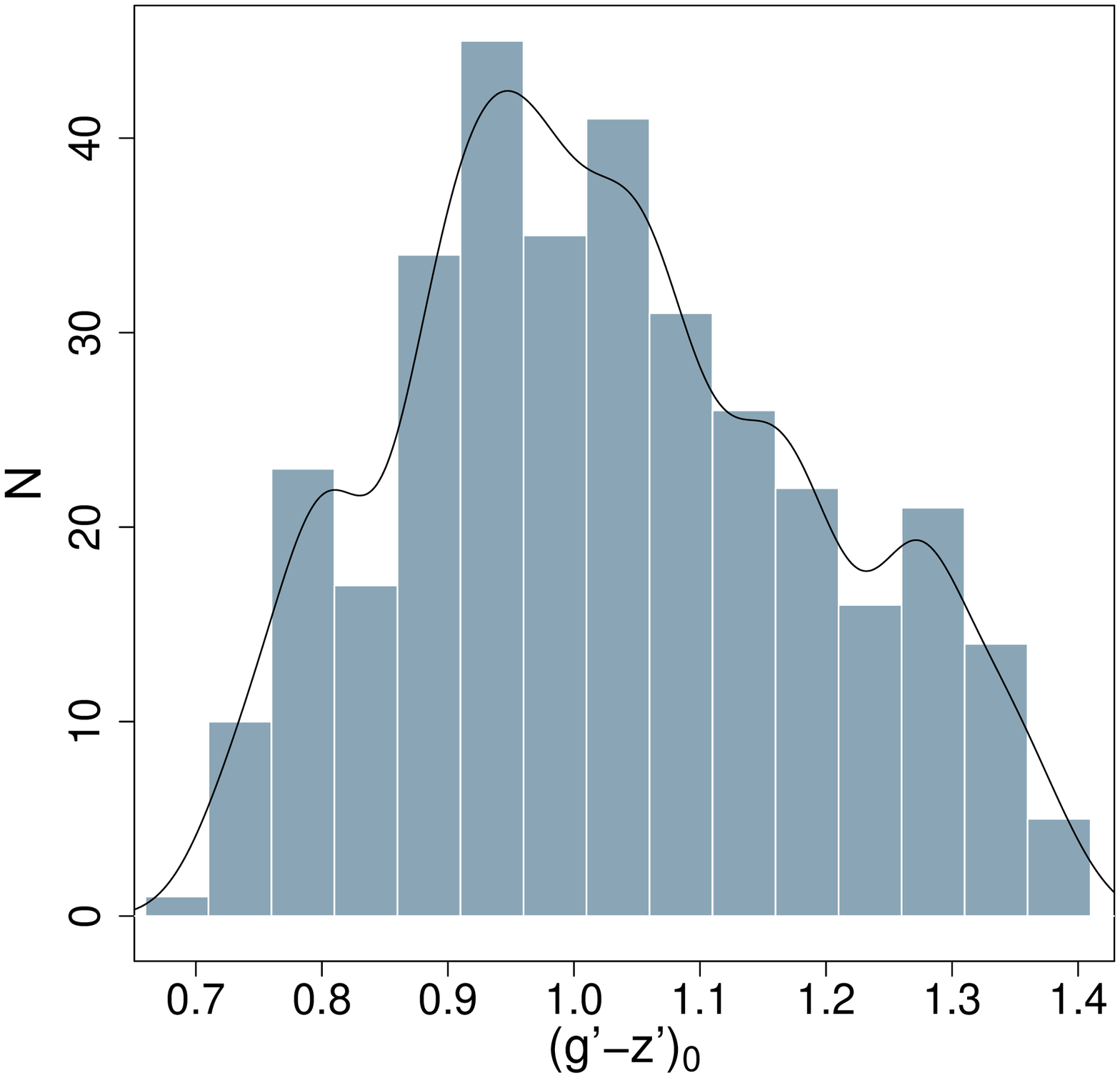}
    \includegraphics[width=0.3\textwidth]{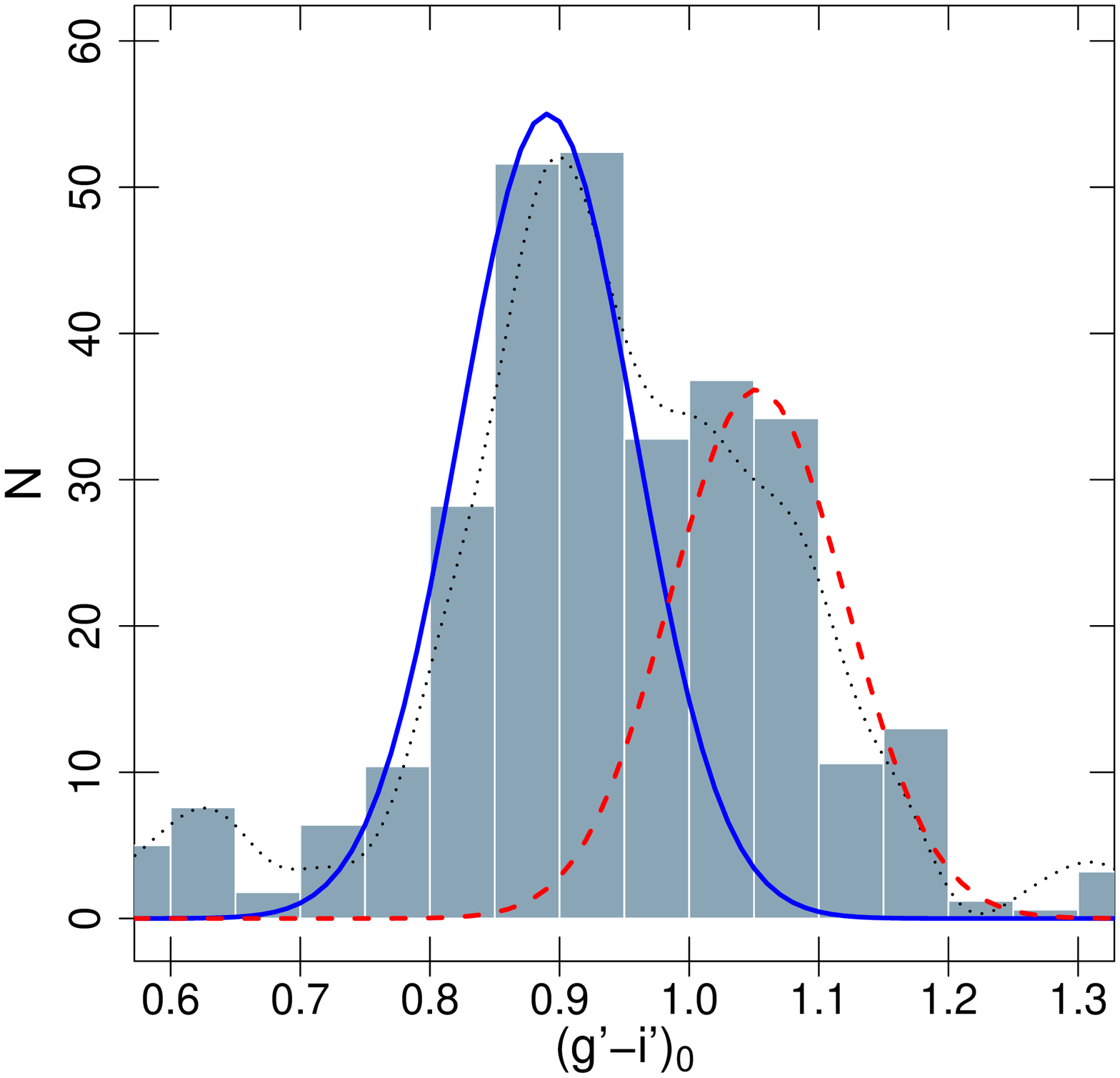}
    \caption{Colour distributions of GC candidates in the GMOS fields. In the top panel, we show the $(g'-z')$ distribution corresponding to the central field, and the solid line indicates the density distribution. In the lower panel, we show the $(g'-i')$ distribution corresponding to the three GMOS fields. The Gaussian fits obtained using a nonlinear least squares function are presented in solid and dashed line, while the dotted line represents the density distribution.}
    \label{fig:dcol1}
\end{figure}

Figure \ref{fig:dcol1} shows two GC candidates colour distributions. The background level was measured using the regions further from the galaxy in the GMOS adjacent fields, and it was deemed negligible. The bottom panel shows the colour distribution for the three GMOS fields in ($g'$-$i'$). Visually, the distribution appears to be bimodal, with a peak near $\sim0.9$ and another one near $\sim1.05$. Using the GMM code developed by \cite{muratov2010}, we obtain these same peaks, but the statistical parameters this code runs return inconclusive results. The first parameter is the kurtosis, which measures the `flatness' of the distribution. A negative kurtosis is considered to be a necessary condition to confirm bimodality, though not sufficient. In this case, the kurtosis is $\sim0.05$.

The other parameter GMM calculates is called $D$, defined as can be seen in Equation \ref{eq:d},  measures the relative distance between the peaks of the subpopulations. In the equation, $\mu_{i}$ and $\sigma_{i}$ represent the mean and the standard deviation calculated for each subpopulation. In a bimodal distribution, $D>2$. For this distribution, $D\sim2$, which, like the kurtosis, is not a conclusive result since it does not support neither unimodality nor bimodality. Looking at the distribution, and specifically focusing on the colour limits, we conclude that compared to GC populations in brighter galaxies, this one is significantly narrower, making it more difficult to separate potential subpopulations through the usual statistical parameters. 

\begin{equation}
D = \frac{|\mu_{1}-\mu_{2}|}{\left[\left(\sigma_{1}^2-\sigma_{2}^2\right)/2\right]^{1/2}}
\label{eq:d}
\end{equation}

Attempts to separate the subpopulations statistically using other clustering methods have also given uncertain results. Instead, we calculated the `skewness' of the distribution, i.e., the estimator for the third momentum. 
We generated several random series of numbers following a normal distribution using the sample's mean and standard deviation, and the GMM values. In both cases, we obtained similar results for the skewness less than 10\,per cent of the time, meaning we can affirm this distribution is not unimodal with 90\,per cent confidence. 

Using the nonlinear least squares function in Rproject \citep{rp}, we were able to fit two Gaussians, which allowed us to estimate the fraction of blue GCs. We obtained that 60\,per cent of the GCs belong to the blue subpopulation. 

The top panel of Figure \ref{fig:dcol1} shows the colour distribution in ($g'$-$z'$) for the central GMOS field, since it is the only one for which we have observations in the $z'$ filter. Comparing our results with the ones presented by \cite{cho2012}, we saw a significant amount of GC candidates in the central region of the galaxy were not being detected in our fields, as a consequence of issues in the observations in the $z'$ filter. In order to make up for this, we used the completeness curves obtained as described in the previous section to estimate what portion of sources we were missing per magnitude bin, and we randomly added the corresponding amount of sources from the photometry performed in the ACS data.

In the resulting distribution, a new feature appears. Between the peaks at $g'-z'=0.96$ and $g'-z'=1.24$ which were established by \cite{cho2012} using KMM, we observe an overdensity which could be an indication of a third peak. A similar result was presented in \cite{escudero2020}, where an intermediate subpopulation is more prominent in $g'-z'$ than in $g'-i'$. 

\begin{figure}
    \centering
    \includegraphics[width=0.22\textwidth]{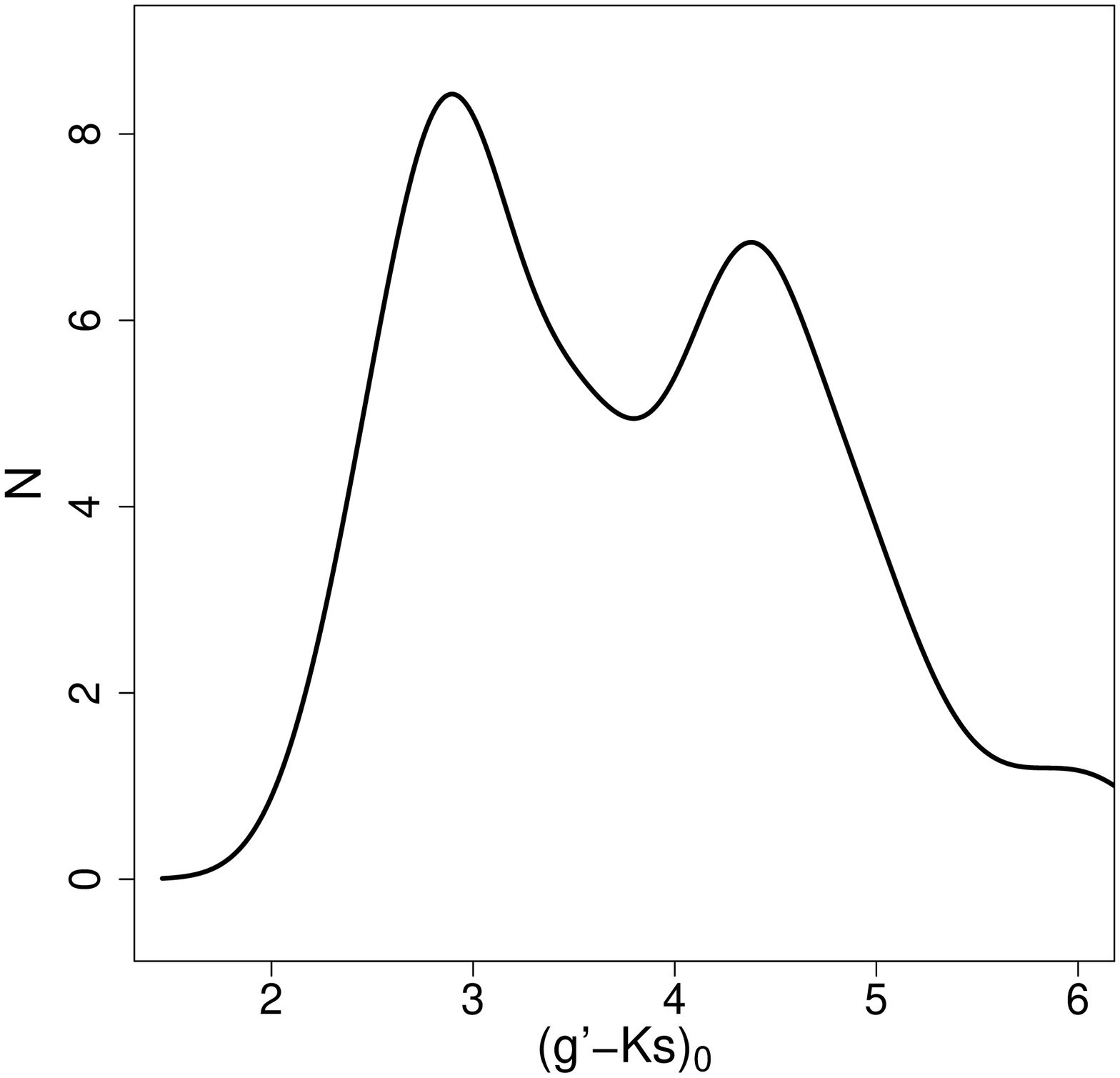}
    \includegraphics[width=0.22\textwidth]{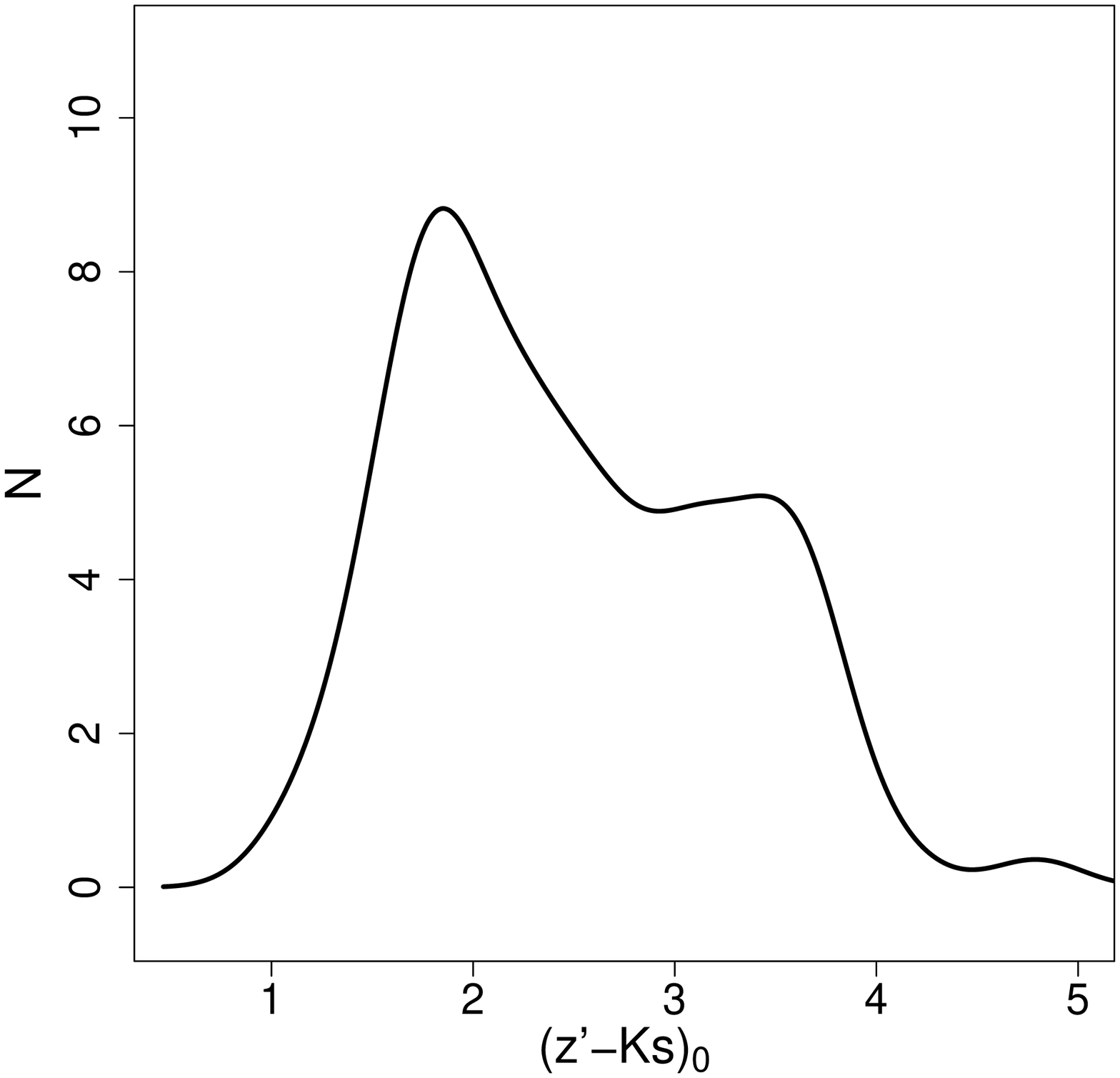}
    \caption{Density curves for the GC colour distributions of the optical-NIR combinations.}
    \label{fig:dcol2}
\end{figure}

In Figure \ref{fig:dcol2}, we show the density colour distributions for two optical-NIR combinations, obtained with a Gaussian kernel. Since the amount of sources was too scarce to obtain a histogram or run statistical tests, we show only the density distribution and limit ourselves to a qualitative analysis. In both cases, two peaks appear to be present, which is consistent with our previous analysis using only optical colors. This supports the presence of two GC subpopulations strongly, since optical-NIR combinations do not always show the same bimodality found in the optical colours \cite{chiessantos2012}. Though the intermediate colours do not show a peak, it is worth noticing that the valleys between the peaks show a difference in skewness that could be consistent with an overdensity in these colours.   

\subsection{Radial distribution}

\begin{figure}
    \centering

    \includegraphics[width=0.45\textwidth]{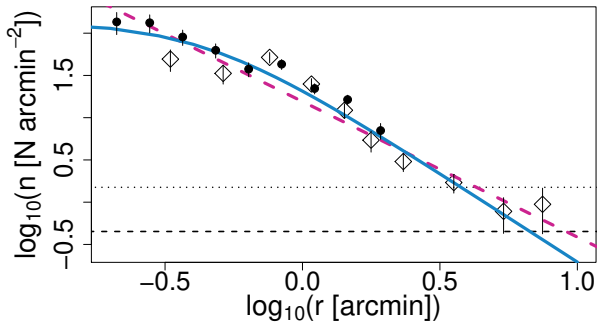}
    \caption{Radial distribution of GC candidates  for the three GMOS fields, shown in black diamonds with error bars. The black circles show the radial distribution for the ACS field. The dotted horizontal line shows the estimated background level, and the horizontal dashed one marks where the 30\,per cent of it is reached. The dashed pink line shows an power-law fit, whereas the solid light-blue line shows a Hubble law fit.}
    \label{fig:drad}
\end{figure}

In Figure \ref{fig:drad} we show the GC radial distribution for the three GMOS fields combined, in addition to the distribution from the ACS field. The Magellan/FourStar images cover a much larger extension, but since we only have one filter for those it is not possible to select GC candidates outside of the GMOS fields. 

The inner region of the distribution is flattened, an effect attributed to the tidal forces being stronger in the regions closer to the centre of the galaxy, which causes lower rates of survival for GCs\citep{kruijssen2015}. It is unlikely that this could be an observational effect, since the completeness correction was carried out taking into account the variations at different galactocentric radius. As a first approximation, we fit a power law, with a slope of $-1.60$. This fit does not take into account the flattening, which is why we also fit a modified version of the Hubble-Reynolds law \citep{binney1987,dirsch2002}, described in Equation \ref{eq:hubble}. In \cite{caso2019b}, the same modified Hubble profile is fitted only to the ACS sample \citep{cho2012}, and the parameters are consistent, as shown in \ref{tab:hub}. 

\begin{equation}
    \centering
    n(r)=a\left(1+\left(\frac{r}{r_{0}}\right)^{2}\right)^{-b}
    \label{eq:hubble}
\end{equation}

\begin{table}
\centering
 \caption{Coefficients 
of the Hubble-Reynolds law obtained from fitting radial distributions. First set of parameters corresponds to this current work, obtained from combining the radial distribution from the GMOS fields with the one from the ACS field. Second set of parameters corresponds to the radial distribution from the ACS field, as seen in \protect\cite{caso2019a}.}
 \label{tab:hub}
 \begin{tabular}{|lll|}
  \hline
 Parameter & GMOS+ACS Values & ACS Values \\
  \hline
 a & $2.15\pm0.14$ & $2.22\pm0.03$\\[2pt]
  \hline
 $r_{0}$ & $0.44\pm0.16$ & $0.51\pm0.06$\\[2pt]
  \hline
 b & $1.05\pm0.14$ & $1.21\pm0.10$\\[2pt]
	\hline
\end{tabular}
\end{table}

We obtained an estimate for the background level using selected regions of the adjacent field, and we approximate the full extension of the GCS as the value at which the Hubble-Reynolds law reaches 30\,per cent of said level. According to the distance to NGC\,1172 adopted in this paper, the full extension is $\sim43$\,kpc. The integration of the Hubble-Reynolds law up to the estimated extension gives us a number for the total population up to $i'=25$\,mag, which in this case is of $400\pm65$ GCs.

We calculated a stellar mass of $\log_{10}(M_{\star}/\rm M_{\odot})=10.18$ using the (M/L) relation based on the $K_s$ magnitude and ($B-V$) colour  \citep{caso2019a}, described by \cite{bell2003}. This allows us to see that the extension of the GCS is consistent with the stellar mass of the galaxy according to the relations obtained in \citep{caso2019a}.

\subsection{Luminosity function}

\begin{figure}
    \centering
    \includegraphics[width=0.45\textwidth]{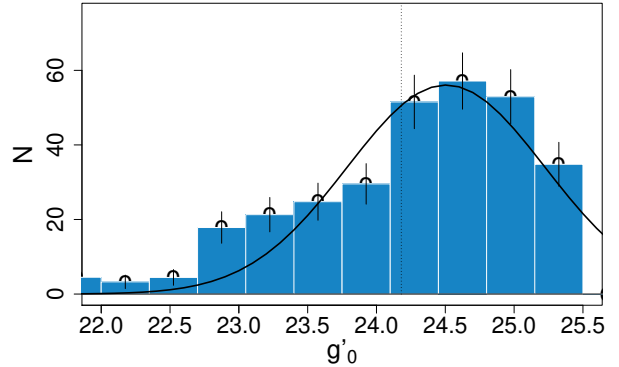}
    \caption{Luminosity function of the GC candidates for the central GMOS field, with the turn-over obtained from \protect\cite{cho2012} shown as a dotted vertical line. The solid black line represents our Gaussian fit.}
    \label{fig:lfum}
\end{figure}

Another important feature of GCSs is the fact that the luminosity function (GCLF), which describes the number of objects per unit magnitude, presents a `universal' turn-over in absolute magnitude \citep[e.g][]{harris2014}, making it useful for estimating distances. \cite{cho2012} derive a distance based on the GCLF which we use as a reference for our results, since they use deep ACS photometry for this purpose. Our independently derived distance modulus is in very good agreement with theirs, as Figure \ref{fig:lfum} shows. A gaussian function with a turn-over magnitude at $g'=24.5$ provides a good representation of the luminosity distribution. We used the $g'$ band for a direct comparison with the results from the literature, and estimated a completeness level of $85$\,per cent at $g'\sim25.5\,mag$.

The luminosity function also allows us to estimate 
what percentage of the GC population we detect on our images. Integrating the Gaussian fit obtained using the parameters from \cite{cho2012} up to our magnitude limit, we calculate our detected population comprises 89\,per cent of the total. Combining this result with one from the integration of the radial distribution, we get a total population of $450\pm72$ GCs. Our own fit to the GCLF results in a total population of $415\pm79$ GCs, which is consistent within the errors.

\begin{figure}
    \centering
    \includegraphics[width=0.45\textwidth]{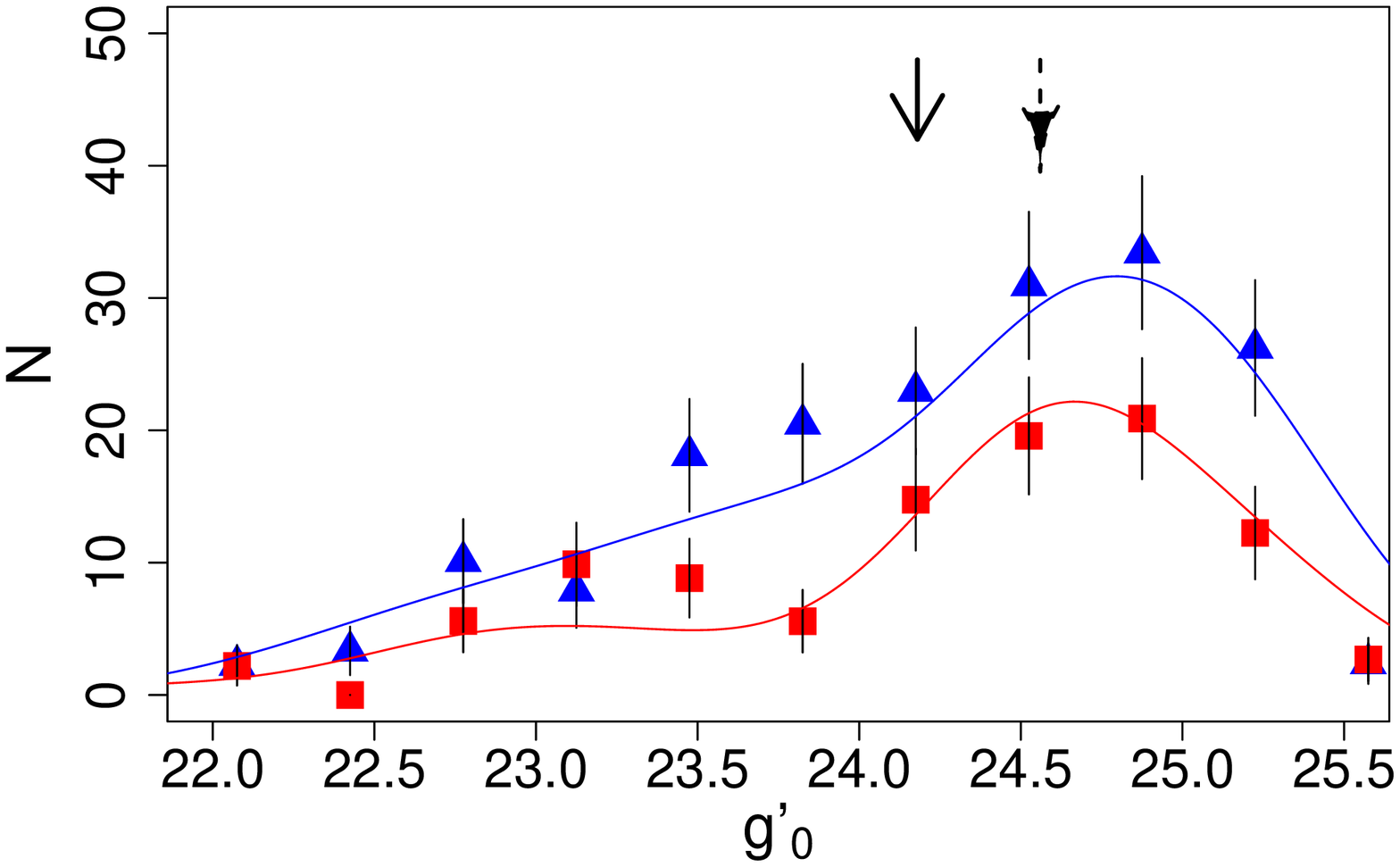}
    \includegraphics[width=0.45\textwidth]{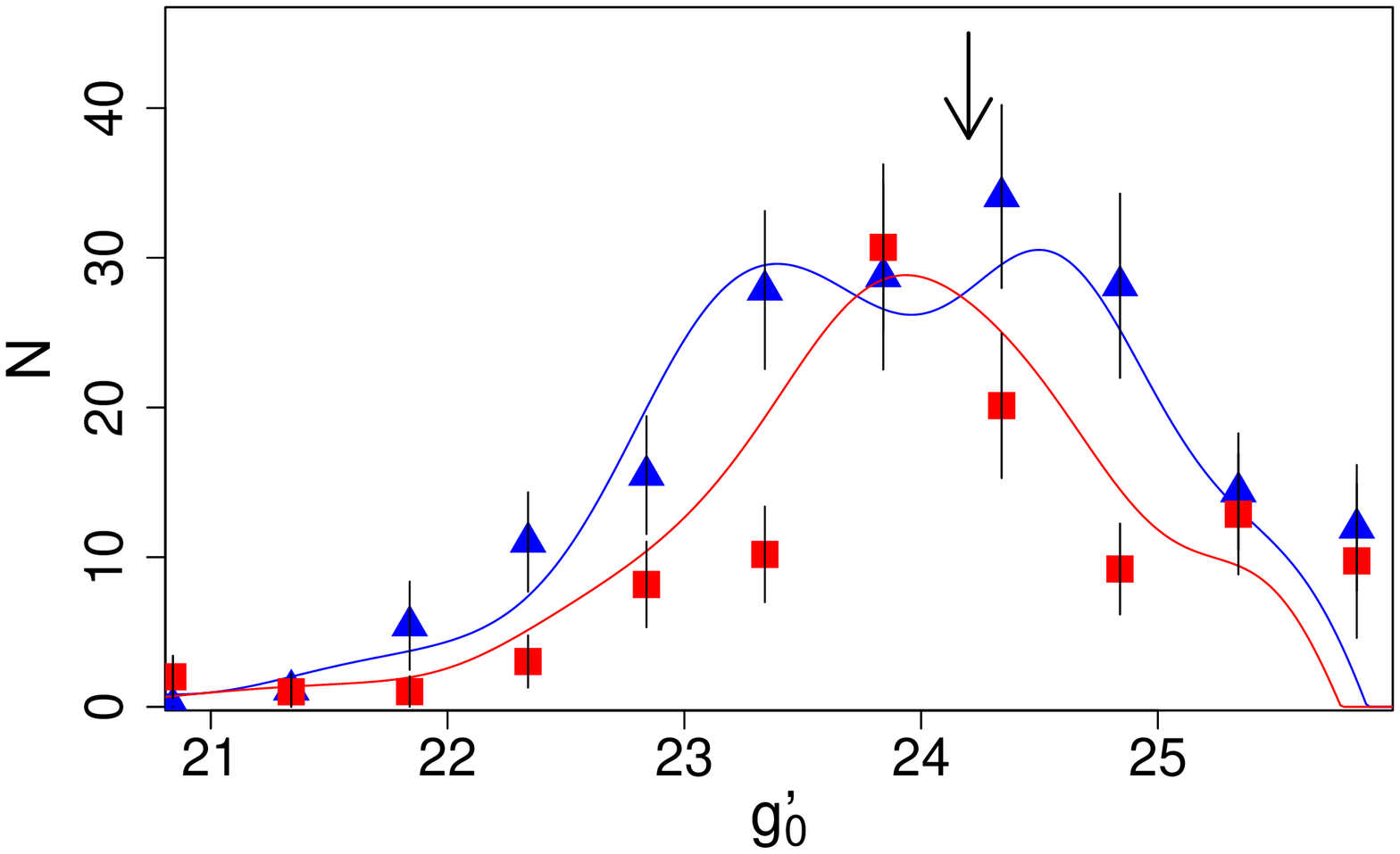}
    \caption{Luminosity functions for the different subpopulations (blue as triangles, red as squares) for the GMOS (top) and the ACS sample (bottom). The solid arrow shows the turn-over from the literature, the dashed arrow shows the one obtained from our fit to the full sample.}
    \label{fig:lfcol}
\end{figure}

 In Figure \ref{fig:lfcol} we show the luminosity functions for the ACS sample and for the GMOS sample, having separated blue and red GCs. Though the distributions present some noise, it can be seen that the turn-over does appear slightly shifted, with the blue subpopulations peaking at a lower magnitude.

\subsection{Specific frequency and total mass}

\begin{figure}
    \centering
    \includegraphics[width=0.9\linewidth]{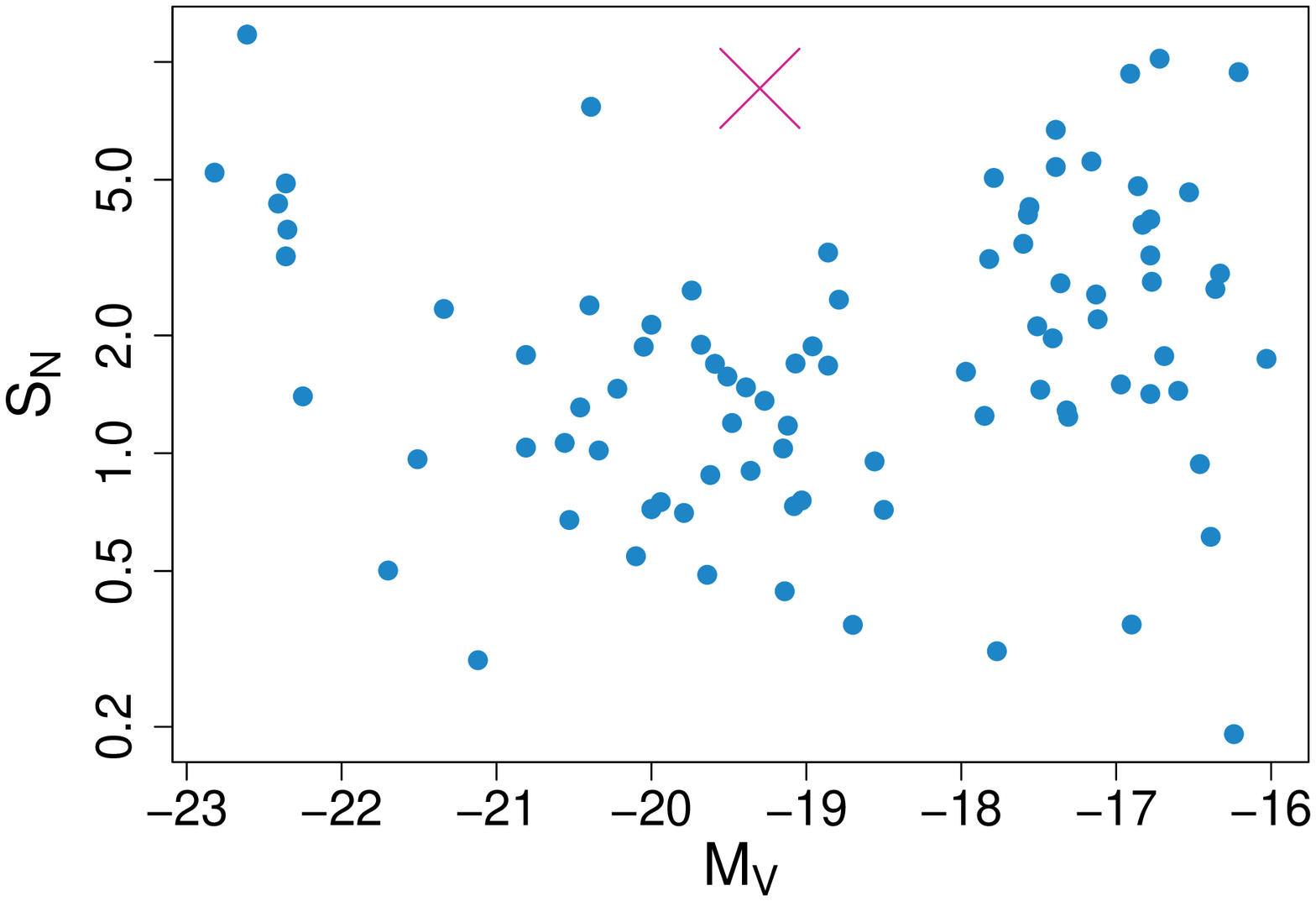}
    \includegraphics[width=0.9\linewidth]{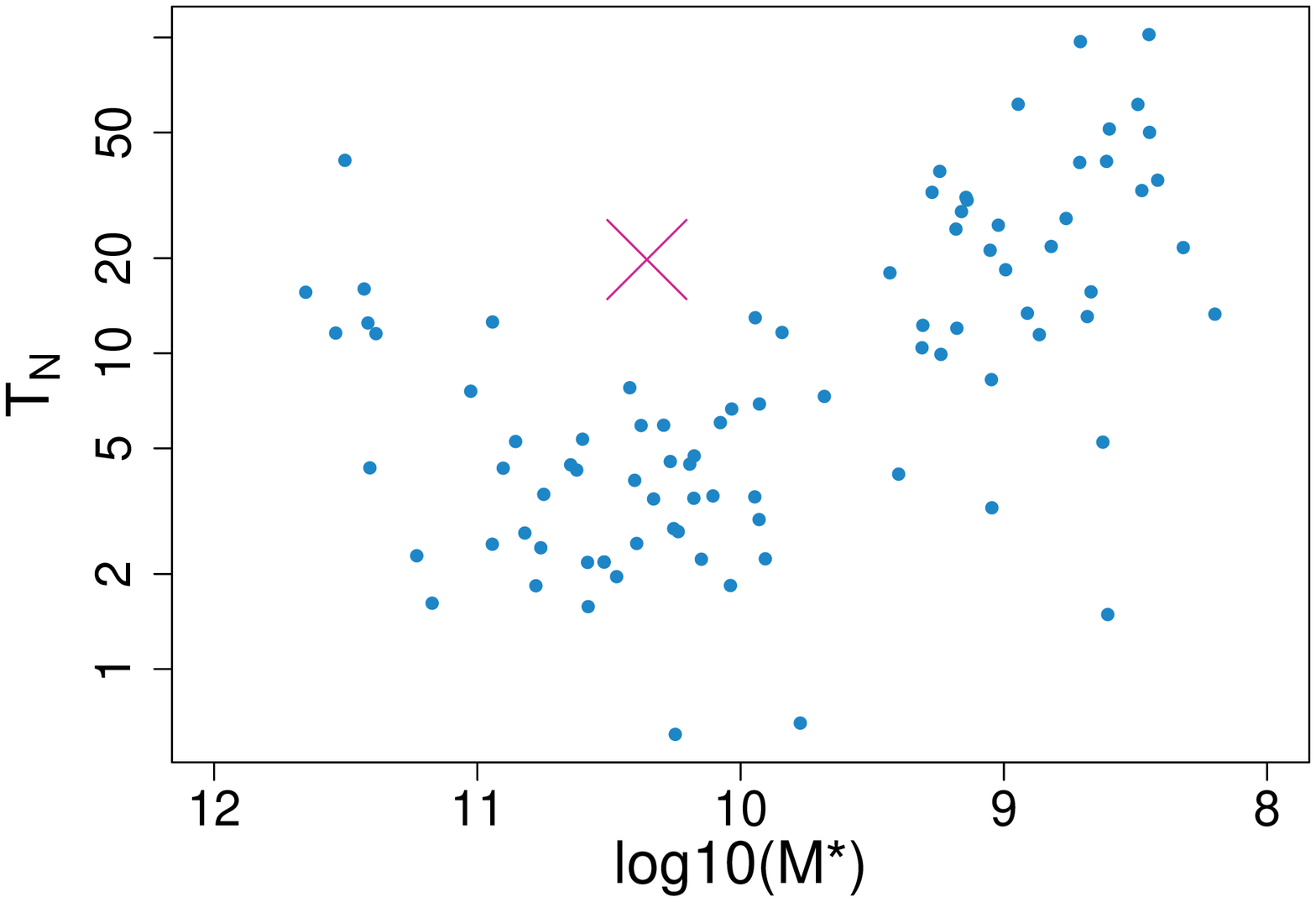}
    \caption{Top panel: Specific frequency in a logaritmic scale versus absolute visual magnitude for a sample of ETGs in the Virgo Cluster shown in light blue dots, and NGC\,1172 shown as a large pink X. Bottom panel: Parameter $T_{N}$ in a logaritmic scale versus the stellar mass for the same sample and NGC\,1172.}
    \label{fig:sntn}
\end{figure}

The specific frequency describes the efficiency of GC formation in comparison to field stars in a galaxy. For NGC\,1172 we obtain a value of $S_{N}=8.6\pm1.5$, which is in agreement with the previous result of $S_{N}=9.18\pm4.41$, presented by \cite{cho2012}. To illustrate the peculiarity of this result, we show in the top panel of Figure \ref{fig:sntn} the specific frequency of a sample of galaxies versus their absolute visual magnitude from \cite{harris2013}. It is easy to see NGC\,1172 is placed on a fairly empty section of the plot, with a specific frequency only comparable with much brighter galaxies. 

Another quantity that reflects the efficiency of GC formation is $T_{N}$, defined by \cite{zepf1993} as the number of GCs per unit of stellar mass. For NGC\,1172 we obtained a value of $T_{N}=19.8$, which is again significantly high for a galaxy of this mass. In the bottom panel of Figure \ref{fig:sntn}, we show the same sample of galaxies used for the specific frequency. We obtained the stellar masses for these galaxies with the method mentioned above, combining the previous catalogue with the magnitudes in \cite{ferrarese2006}. 

Another quantity we can obtain based on the number of GCs is the total mass of the GCS. If we consider the mean mass of a GC to be $M_{\rm GC}\sim3.25\times10^{4}\rm{M}_{\odot}$ \citep{harris2017b}, we obtain an estimate of the total mass of the GCS of $1.5\times10^{7}\rm{M}_{\odot}$. 


\subsection{Photometric metallicities}

\begin{figure*}
    \centering
    \includegraphics[width=.3\textwidth]{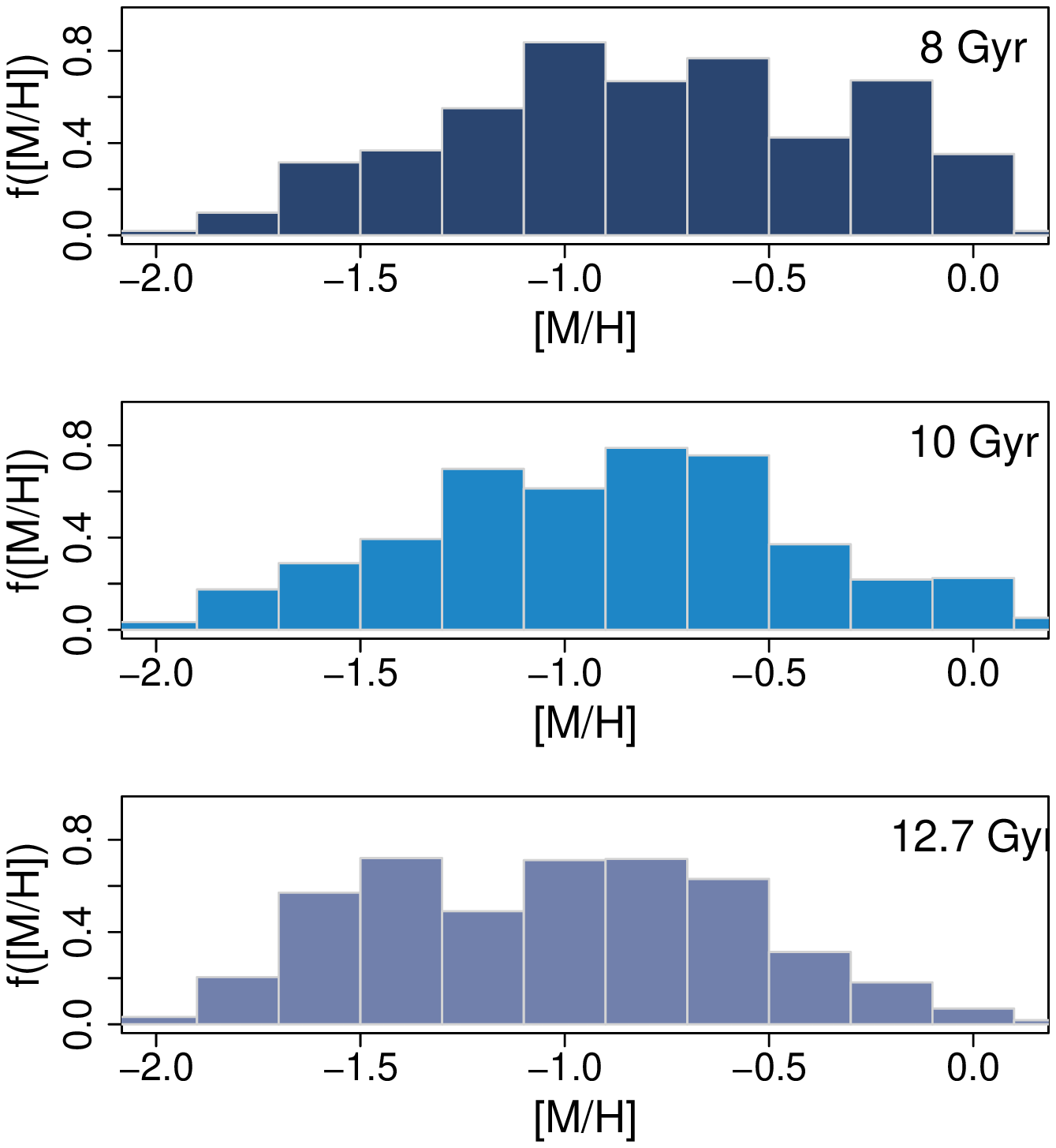}
    \includegraphics[width=.3\textwidth]{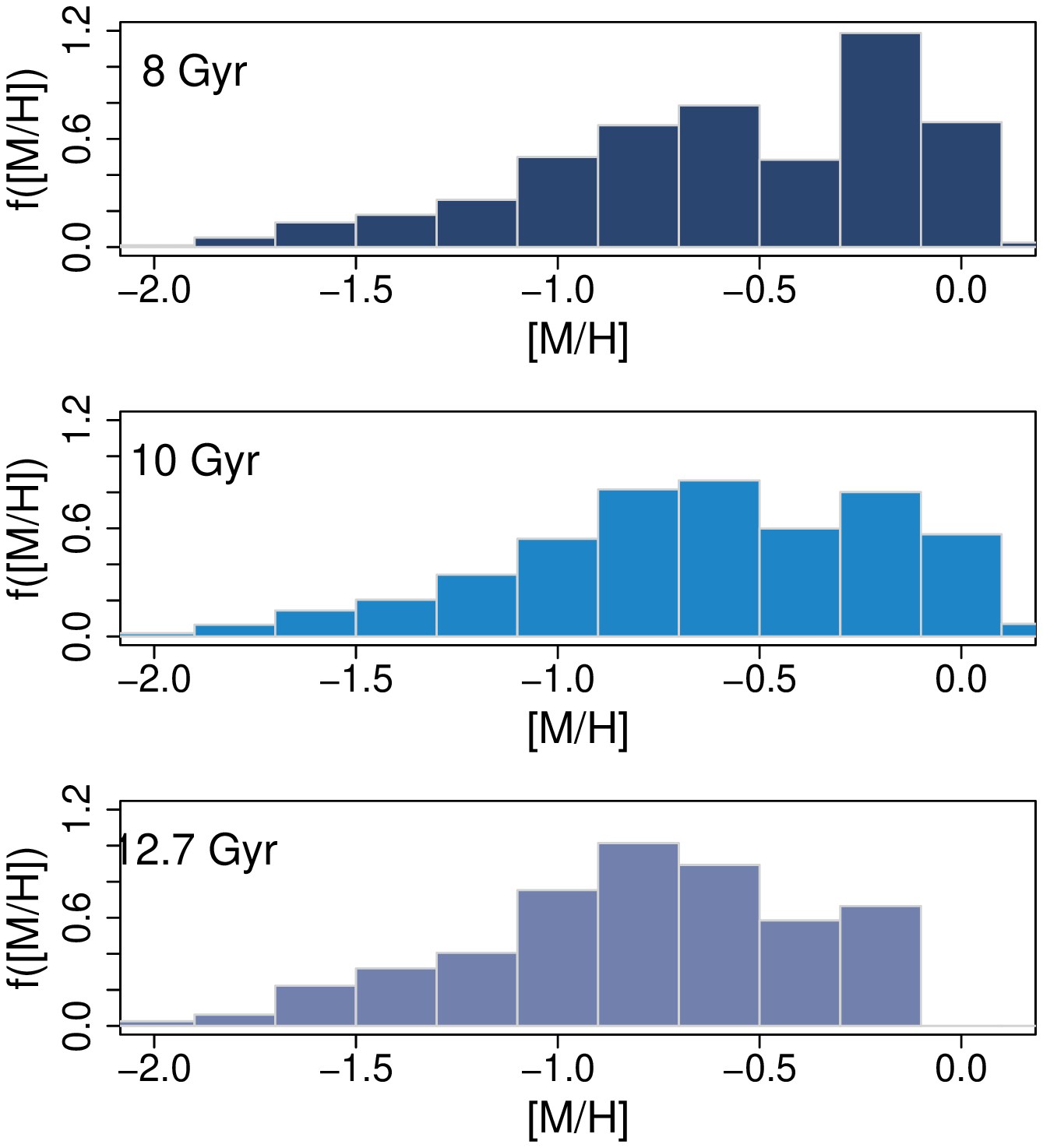}
    \includegraphics[width=.3\textwidth]{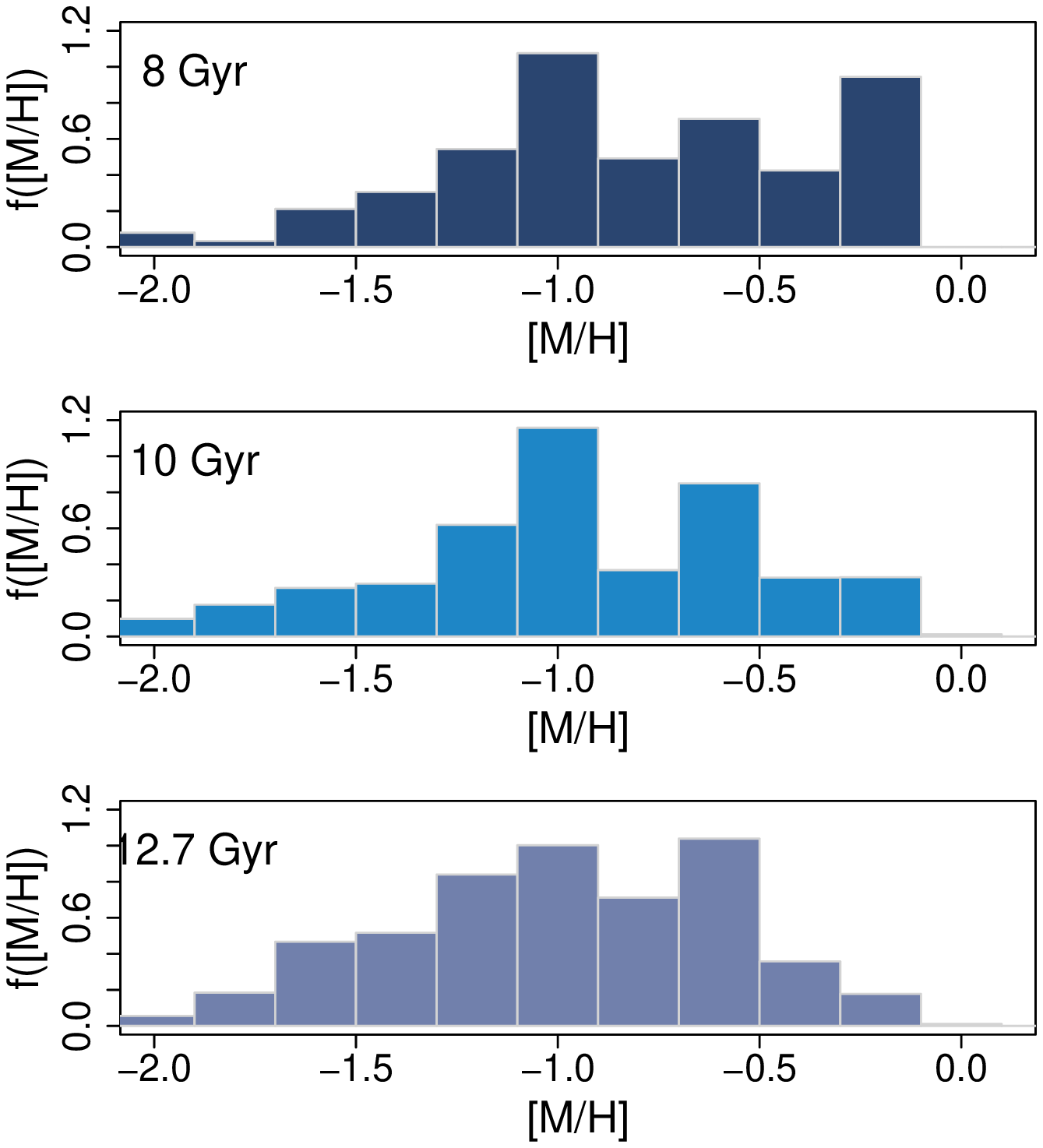}
    \caption{Probability function for the metallicity obtained for different ages, using different colour combinations. Left panel shows Sample A (corresponding to the central GMOS field, uses $g'r'i'z'$ magnitudes), middle panel shows Sample B (corresponding to the three GMOS fields, but only for the bright fraction of objects for which $Ks$ magnitudes were available, uses $g'r'i'Ks$ magnitudes), and right panel shows Sample C (corresponding to the central panel, for objects for which both $z'$ and $Ks$ magnitudes were available, uses $g'r'i'z'Ks$ magnitudes). From top to bottom, ages increase, and the metallicites move towards lower values. There are no significant differences between the samples that are not consistent with this shift, and in all cases, they point to the underlying distribution being bimodal.}
    \label{fig:dmet}
\end{figure*}

As was mentioned above, colour distributions may not be an accurate representation of the metallicity distribution of a GCS. However, the combination of optical and near-infrared colours has been shown to be a more precise representation \citep[e.g.][]{hempel2007,tudorica2015}. Therefore, we combined different colours and obtained photometric metallicities in order to compare them and obtain an estimate of the metallicity distribution, in a similar way as \cite{forte2013} and \cite{caso2017}.

In order to do this, we adopted a Bayesian approach. Bayes' theorem can be expressed as seen in Equation \ref{eq:bayes}, where $P(A|B)$ in this case stands for the probability of obtaining a certain metallicity distribution (A) given our colour distribution (B), and it is proportional to the product of the likelihood function and the prior probability. 

\begin{equation}
    P(A|B) \propto P(B|A)P(A)
    \label{eq:bayes}
\end{equation}

\begin{figure}
    \centering
    \includegraphics[width=.45\textwidth]{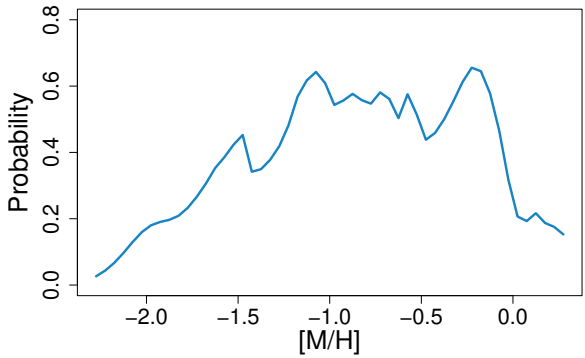}
    \caption{Smoothed metallicity distribution obtained for the spectroscopic data, used as prior distribution.}
    \label{fig:apriori}
\end{figure}

From \cite{fahrioncat} we obtained a catalogue of GCs located in the Fornax cluster, with metallicities measured using the Multi Unit Explorer Spectrograph (MUSE) mounted on the VLT. This sample also included $g'$ and $z'$ magnitudes from \cite{jordancat}, which were used to randomly substract GCs in order to build a colour distribution with a closer resemblance to the $(g'-z')_{0}$ colour distribution from NGC\,1172. The metallicity distribution from the resulting sample was built assuming that each GC presents a normal metallicity distribution, with mean equal to the spectroscopic measurement and uncertatinty equal to the dispersion. This smoothed metallicity is assumed as the prior probability, and is shown in Figure \ref{fig:apriori}.

 In order to build the likelihood function, first we generated integrated magnitudes using simple stellar population (SSP) models, making use of the CMD $3.1$ web interface\footnote{\url{http://stev.oapd.inaf.it/cgi-bin/cmd_3.1}} which computes isochrones and derivatives. We used the PARSEC version 1.2s evolutionary tracks and the log-normal initial mass function from \cite{chabrier2001}, and generated sequences of isochrones of constant age, with the metallicity (Z) covering a range between  $[M/H]=-2.1$ and $0.03$, with a step of $0.001$. Considering GCs are thought to have similar ages, generally around $10$\,Gyr, we did this for three different constant ages, $8$, $10$, and $12.7$\,Gyr, to compare and look for any significant differences between them. 

According to the different FOV for each filter, we built three samples using the colours available for GC candidates in the present work. Sample A corresponds to the central GMOS field and uses magnitudes $griz$, combined as  $(g'-i')_{0}$ and $(g'-z')_{0}$. For a fraction of bright GCs across the three GMOS fields, $g'r'i'Ks$ photometry was also available, and so Sample B combined $(g'-i')_{0}$ and $(g'-Ks)_{0}$. Sample C covers only the central GMOS field for which both $Ks$ and $z'$ photometry are available simultaneously, combined as $(g'-i')_{0}$, $(g'-z')_{0}$ and $(r'-Ks)_{0}$. These last two samples have a limited size due to the lack of depth in the $Ks$ filter, but are interesting to analyse since they cover the largest colour range. For each GC, its set of colours was assumed as a normal distribution centred on the colours resulting from their measured magnitudes, with the corresponding dispersions derived from the uncertainties. We built a grid of colours spanning up to $3\,\sigma$ from the mean colours, and calculated an estimate of metallicity for each of them by interpolating theoretical metallicities, looking for the minimum distance to the SSP curve in the colour space. Finally, the probability for each cell from the colour distribution function was used to build the likelihood distribution, following the limits and binning of the prior distribution. 

The final metallicity probability for each GC was then obtained multiplying the likelihood function and the prior probability. We then added all of the probabilities within each sample in each metallicity bin, and obtained metallicity distributions for the GCS which are shown in Figure \ref{fig:dmet}. Though the distribution including the largest amount of filters (Sample C) appears to be noisier due to the smaller amount of GCs, the distributions for each colour sample do not show significant differences between them. In the case of the variations with age, we can see that as we go from the youngest to the oldest sample, the metallicities shift towards lower values, as expected.

\begin{figure}
    \centering
    \includegraphics[width=.4\textwidth]{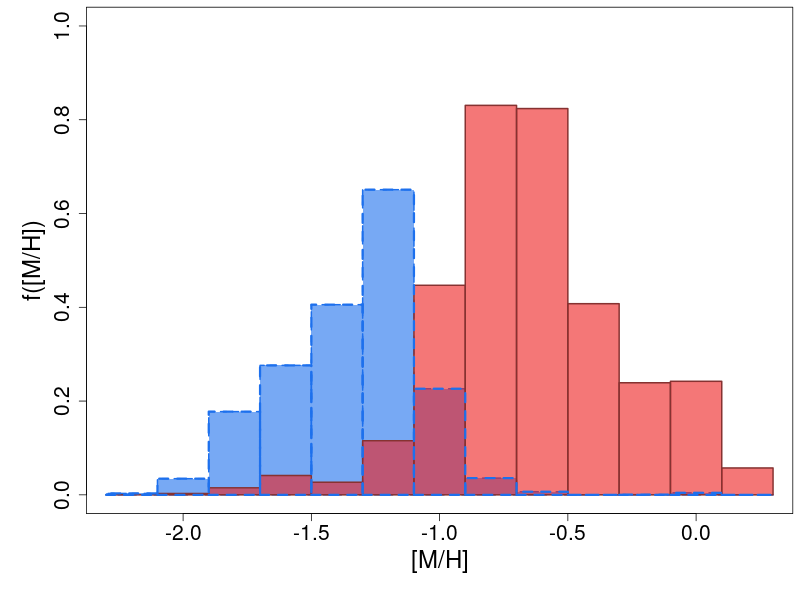}
    \caption{Probability function of the metallicity distributions obtained using $(g'-i')_{0}$ and $(g'-z')_{0}$ for 10 Gyr for the red (solid lines) and blue (dashed lines) subpopulations.}
    \label{fig:colmet}
\end{figure}

In addition, we show in Figure \ref{fig:colmet} the metallicity distribution for Sample A, which has the largest amount of GCs, using an age of 10\,Gyr, separating blue and red GCs at the limiting value of $(g'-i')_{0}=0.95$\,mag. The separation remains clear in metallicity, which supports the presence of the two GC subpopulations.

\section{Discussion}

There are a number of processes involved in the assembly of a GCS that can influence the number of members, mainly the formation of the GCS, accretion, and its tidal destruction of the less massive members. If the high number of GCs were a consequence of a distinctly high {\it in situ} GC formation efficiency, the red GCs fraction would be expected to be significantly higher, like the inner GCS in many massive ellipticals \citep[e.g.][]{peng2006,kartha2016,caso2017}. In the case of NGC\,1172, we do not find a significantly large red subpopulation. Another possibility is NGC\,1172 having formed GCs with the typical rate for galaxies with a similar mass, but then being less efficient at destroying them. An important issue to take into account in this case is that a lower efficiency in disruption processes could affect the mass function of the GCS (GCMF). Although the treatment of physical processes vary in numerical studies, in recent years several have pointed to tidal stripping and tidal shocks as the main processes at work in GCs disruption  \citep{kruijssen2015,lamers2017,li2019}. These would be relevant in shaping the mass function of GCs, because tidal processes are more effective in less massive GCs, and lower efficiencies would then lead to an excess of less massive GCs in the GCMF, similar to observational results in some bright merger remnants \citep[e.g.][]{goudfrooij2007} and nearby dwarf galaxies \citep[e.g.][]{sharina2006}.
In our luminosity function and in the one presented by \cite{cho2012}, a Gaussian fit is statistically closer to the shape of the distribution, with a dispersion in the range of expected values for the stellar mass of the galaxy \citep[e.g.][]{jordan2007,masters2010}. In both cases, the turn over magnitude is in agreement with the usual for old GCs in bright early-type galaxies \citep{jordan2004,harris2014}, considering the distance to NGC\,1172 indicated in Section\,\ref{intro.sec}. It seems unlikely then that enough low-mass GCs have survived to shift the number of the population so significantly. The colour distribution, with a large fraction of GCs in the range of typical metal-poor ones, also disfavours this scenario. It is relevant to this last point to trace a comparison to the study of isolated ellipticals presented by \cite{salinas2015}, where most of the analysed GCS present significantly predominant blue subpopulations in a much larger fraction than in this case. In addition, there is the opposite scenario presented in the study of NGC\,1277, a galaxy described as candidate for being a relic elliptical, without a significant mass contribution due to merging processes. The GCS of NGC\,1277 presents a unimodal colour distribution, lacking of a metal-poor subpopulation \citep{beasley2018}. NGC\,1172 does not reach either of these extremes, instead showing a behaviour similar to most early-type galaxies found in more crowded environments.

The scaling relations of GCS in nearby galaxies show large values of specific frequencies ($S_N$) and  $T_N$ parameters at both ends of the stellar mass distribution, with a minimum reached for galaxies presenting intermediate masses, due to a combination of external and internal processes that affect both the formation efficiency of GCs and the stellar field population \citep{harris2013}. Interestingly, \cite{georgiev2010} indicated this minimum at a magnitude of $M_{V} \approx -20.5\,mag$ ($L_V \approx 2\times 10^{10}\,{\rm L_{\odot}}$), which does not differ significantly from NGC\,1172 luminosity. In the particular case of massive ellipticals in high density environments, which usually present extremely populated systems \citep[e.g.][]{peng2011,harris2017a}, these large parameters are also related with the accretion history of the galaxy, responsible for a considerable fraction of its GCs \citep{forbes2011}, which leads to extended GCS up to several tens of kiloparsecs, and radial gradients in the colour peaks of the GCs subpopulations \citep[e.g.][]{bassino2006,caso2017,debortoli2020}. 

In the case of NGC\,1172, the environment as seen in the present day does not appear to support this hypothesis at first glance,  since there are no close neighbours. However, it could have had neighbours in its early days and the high population of GCs could still be a product of accretion. The large fraction of blue GCs suggests that the accretion of satellites played a main role in the evolutionary history of NGC\,1172, although the lack of tidal features present in many field ellipticals \citep{tal2009} disfavours late mergers. Considering the relation derived by \cite{hudson2014,harris2017b} for the halo mass and the mass enclosed in GCs, the total mass of NGC\,1172 is $6.3\times10^{11}\rm{M}_{\odot}$, pointing to a dark matter dominated halo. As mentioned before, some dwarf galaxies also present high values of $S_{N}$ and $T_{N}$, due to their accretion providing GCs without increasing the stellar mass of the host galaxy \citep{georgiev2010,harris2013}.

The other peculiarity the GCS of NGC\,1172 shows is its colour distribution, which deviates from unimodality but cannot be easily divided into subpopulations, due to being too narrow and having a minimal distance between peaks. However, the estimator of the third momentum (skewness) of the colour distribution derived in Section\,\ref{dcol.sec} deviates from the typical values for samples of similar size generated by a single Gaussian distribution with a 90\,per cent of confidence, pointing to a bimodal distribution. Several nearby galaxies have been pointed as unimodal distributions, but later analysis has proven they present several peaks in their colour distribution (e.g., the elliptical NGC\,4365, \citealt{kundu2005,blom2012}, the  merger remnant NGC\,1316, \citealt{richtler2012,sesto2016}), or even presenting variations in the same GCS as a function of the galactocentric distance  \citep{ko2019}. In the case of NGC\,1172, the smaller population of its GCS makes the analysis more difficult. \cite{cho2012} showed that the colour peak of the red GCs strongly depends on the galaxy stellar mass, with blue ones presenting a less clear trend. A similar result is derived from \cite{peng2006} survey of GCS in the Virgo cluster. From these studies, it is expected for intermediate mass galaxies like NGC\,1172 to present smaller distances between the two peaks. On the other side, \cite{lee2019} have pointed that the colour distribution of the GCS from the majority of the galaxies in their sample, could be explained due to a non-linear relation between metallicity and colour. This scenario has been previously stated in the literature \citep[e.g.][]{richtler2006}, but differences in the properties of blue and red subpopulations, including spatial distributions \citep[e.g.][]{bassino2017} and kinematic behaviour \citep[e.g.][]{schuberth2010}, support the existence of (at least) two GC subpopulations. Regarding the presence of a bimodal GC  distribution, results in literature might differ between galaxies \citep{chiessantos2012} and further work is needed on both the observational and theoretical framework \citep{powalka2017}. In the case of NGC\,1172, the photometric metallicities clearly point to a bimodal distribution. 

The most supported theory of formation of GCS currently points to most of the more metallic GCs having been formed in-situ, while a large portion of the less metallic ones are accreted from satellite dwarf galaxies \citep{forbes2011,kruijssen2019}. The differences found in NGC\,1172 when we analyse the luminosity function of each subpopulation separately support this theory, since the slight shifts in the turn-over magnitude for each are thought to be associated with galaxies with lower mass presenting fainter turn-overs \citep{villegas2010}. It is consistent then that less-metallic GCs, which would have formed in low-mass galaxies, present fainter turn-overs.

In this context, the colour distributions all support the hypothesis mentioned above of NGC\,1172 having accreted a significant fraction of its GCS. In comparison with the mean colours for each subpopulation of other ETGs, NGC\,1172 presents a slightly redder blue subpopulation, and a shift in the red subpopulation towards the blue. These shifts could be the cause of the intermediate peak seen in some colours, since the presence of a third subpopulation with intermediate metallicites seems unlikely considering no distinct properties of the GCs in that colour range show when examining their spatial distribution or their position in the CMD.

Being mostly formed in-situ during major star formation episodes \citep{kruijssen2015,choksi2019}, 'red' GCs have imprinted the chemical composition of the gas supply that triggered their formation. The field stellar population of nearby galaxies usually presents a larger metallicity than found for old GCs, due to a more widely spread process of stellar formation in time \citep{guglielmo2015,usher2019}, hence its values could be considered as upper limits. In the case of NGC\,1172, its metallicity has not been studied in detail, but there is enough data to explore its colour. It has a magnitude of $M_{g}=-19.58$\,mag considering the distance adopted in this paper, and from \cite{caso2019b} we obtain $(g'-z')_{0}=1.22$\,mag. \cite{ferrarese2006} analysed the colour-magnitude relation for a significant number of ETGs in the Virgo Cluster and, using this relation, we obtain an expected value of $(g'-z')_{0}=1.45$\,mag for NGC\,1172. This points to NGC\,1172 being bluer than expected, which would justify its red GCs being shifted to the less metallic side. This is in agreement with previous studies focused on field ellipticals, which found bluer colours in comparison with cluster counterparts of similar stellar mas \citep[e.g.][]{lacerna2016}.


\section{Conclusions}
We have presented the photometric analysis of the highly populated GCS of the field elliptical NGC\,1172, using data obtained from GMOS at Gemini, FourStar at Magellan, and archival data from ACS at \textit{HST}. Our principal findings are:
\begin{itemize}
    \item The colour distribution of the GCS is narrow and does not show a clear separation between subpopulations. However, it deviates strongly from unimodality, as can be seen in the skewness of the distribution. The apparent low metallicity of the galaxy could account for the narrowness of the distribution, making the red GCs, formed mainly in-situ, to be slightly bluer than expected. The intermediate overdensity seen in $(g'-z')$ does not appear as clearly in any other colours, but it does not disappear either. However, we find no peculiarities in its spatial distribution or its range of magnitudes that separate it from the rest of the GCS.
    \item The global radial distribution matches the one obtained using the central field only, and from it we obtain an extension of 43\,kpc for the GCS, and a total population up to our completeness level ($i'=25$\,mag) of $400\pm65$ GCs. 
    \item The luminosity function is described by a Gaussian, and integrating it gives us a total population of $450\pm72$ GCs.
    \item The specific frequency $S_{N}=8.6\pm1.5$ and the parameter 
    $T_{N}=19.8$ are significantly high for a galaxy of intermediate luminosity like NGC\,1172. The fact that the luminosity function follows a Gaussian distribution indicates that a low efficiency at tidal destruction is not the cause, while the high fraction of blue GCs does not support the case of a high in-situ formation scenario. The possibility that remains is that the environment of NGC\,1172 presented neighbors which were accreted early on.
\end{itemize}

\section*{Acknowledgements}

The authors would like to thank the referee for providing constructive comments which contributed to improving the paper. We also thank Mischa Schirmer for his helpful feedback on THELI, as well as  Maren Hempel for her advice and guidance. This work was funded with grants from Consejo Nacional de Investigaciones   
Cient\'{\i}ficas y T\'ecnicas de la Rep\'ublica Argentina, Agencia Nacional de Promoci\'on Cient\'{\i}fica y Tecnol\'ogica, and Universidad Nacional de La Plata, Argentina. \\
Based on observations obtained at the Gemini Observatory (GS-2016B-Q37 and GS-2017B-Q38), which is operated by the Association of Universities for Research in Astronomy, Inc., under a cooperative agreement with the NSF on behalf of the Gemini partnership: the National Science Foundation (United States), the National Research Council (Canada), CONICYT (Chile), the Australian Research Council (Australia), Minist\'{e}rio da Ci\^{e}ncia, Tecnologia e Inova\c{c}\~{a}o (Brazil) and Ministerio de Ciencia, Tecnolog\'{i}a e Innovaci\'{o}n Productiva (Argentina). 
This paper includes data gathered with the 6.5 meter Magellan Telescopes located at Las Campanas Observatory, Chile.
This research has made use of the NASA/IPAC Extragalactic Database (NED) which 
is operated by the Jet Propulsion Laboratory, California Institute of Technology, 
under contract with the National Aeronautics and Space Administration. This publication makes use of data products from the Two Micron All Sky Survey, which is a joint project of the University of Massachusetts and the Infrared Processing and Analysis Center/California Institute of Technology, funded by the National Aeronautics and Space Administration and the National Science Foundation.

\section*{Data Availability}
All raw data from GMOS/Gemini and ACS/HST can be found in their respective archives. FourStar/Magellan raw data is available upon request.




\bibliographystyle{mnras}
\bibliography{biblio} 



\bsp	
\label{lastpage}
\end{document}